# A Multi-User Perspective for Personalized Email Communities

Waqas Nawaz[a], Kifayat-Ullah Khan[b], Young-Koo Lee[b,*]

[a]*Institute of Information Systems, Innopolis University, Russia*
[b]*Department of Computer Engineering, Kyung Hee University, Korea*

## Abstract

Email classification and prioritization expert systems have the potential to automatically group emails and users as communities based on their communication patterns, which is one of the most tedious tasks. The exchange of emails among users along with the time and content information determine the pattern of communication. The intelligent systems extract these patterns from an email corpus of single or all users and are limited to statistical analysis. However, the email information revealed in those methods is either constricted or widespread, i.e. single or all users respectively, which limits the usability of the resultant communities. In contrast to extreme views of the email information, we relax the aforementioned restrictions by considering a subset of all users as multi-user information in an incremental way to extend the personalization concept. Accordingly, we propose a multi-user personalized email community detection method to discover the groupings of email users based on their structural and semantic intimacy. We construct a social graph using multi-user personalized emails. Subsequently, the social graph is uniquely leveraged with expedient attributes, such as semantics, to identify user communities through collaborative similarity measure. The multi-user personalized communities, which are evaluated through different quality measures, enable the email systems to filter spam or malicious emails and suggest contacts while composing emails. The experimental results over two randomly selected users from email network, as constrained information, unveil partial interaction among 80% email users with 14% search space reduction where we notice 25% improvement in the clustering coefficient.

*Corresponding author. Young-Koo Lee (yklee@khu.ac.kr), Tel: +82-31-201-3732
*Email addresses:* w.nawaz@innopolis.ru (Waqas Nawaz), kualizai@khu.ac.kr (Kifayat-Ullah Khan), yklee@khu.ac.kr (Young-Koo Lee)

## 1. Introduction

Email has become one of the most imperative asynchronous human communications. Many email systems,namely Gmail, Hotmail and Yahoo-mail, provide services to exchange information among users. Recent statistics show that 4087 million email accounts exist and 200 billion emails are exchanged worldwide each day (Radicati & Hoang, 2011). It is non-trivial for email systems to provide customized services to users based on the huge amounts of electronic data (Dabbish & Kraut, 2006) (Wattenberg et al., 2005). Therefore, personalized information is utilized for this purpose, e.g. Gmail introduces a generic filter to group emails automatically into five categories (Gaikar, 2013). Our focus is to assist these systems for providing personalized services such as email strainer, malicious email identification, email group prediction, contact suggestions while composing emails, and guilt by association.

The analysis of email data[1] is different from other off-line social network analysis (Johnson et al., 2012) (Qu et al., 2014) (Liu et al., 2015). Community detection (Fortunato, 2010), as a major topic in network analysis, has received a great deal of attention with the knowledge of entire email network (Moradi et al., 2012) (Liu et al., 2007). Discovering inherent community structures can help us understand the email network more deeply and reveal interesting properties shared by the members. People belonging to the same community are expected to have similar communication behaviors. Therefore, the identified communities can be used for classifying emails, discovery of prominent users, and highlighting abnormal activities inside the network (Shetty & Adibi, 2005) (Yang et al., 2010) (Wilson & Banzhaf, 2009) (Liu et al., 2009) (Martin et al., 2005) (Nagwani & Bhansali, 2010) (Johansen et al., 2007) (Yoo et al., 2009) (Lin, 2010) (Yelupula & Ramaswamy, 2008) (Timofieiev et al., 2008) (Tan et al., 2014).

It is essential to utilize structural and contextual network information to determine the communication behavior of the users. In structural approaches, the relationships among individuals based on their interactions are analyzed in an email network, which is a kind of social network. In order to analyze the structure of email networks, several Social Network Analysis (SNA) techniques are adopted, such as node centrality (Freeman, 1978), cluster analysis (Clauset et al., 2004) (Papadopoulos et al., 2012), and topological structure (Ahuja, 2000) (Zhu et al., 2013). This kind of analysis reflects either similar neighborhood structures for email communications, e.g., frequent email exchanges with shared neighbors, or communication intensity. The unification of structural and semantic information is also achieved for community analysis (Zhao et al., 2012) (Liu et al., 2007).

Recently, P. Oscar Boykin et al. and Shinjae Yoo et al. claim that a global social network may include noisy features and de-emphasize personalization in the inductive learning of important features through the network. It may also affect the user's own communication behavior or pattern. Consequently, the concepts of localization(Hu & Lau, 2012) or personalization (Boykin & Roychowdhury, 2004) (Yoo et al., 2009) are introduced in literature. Inherently, both local and personalized approaches utilize the personal information owned by a particular user, e.g., all emails from single account. In other words, the personalization concept limits the topological information of the email network up to two-hops from a reference user to identify the communities. Fig. 1, shows the categorical views of the network information where each node and link represent a user and email exchange respectively. The global view refers to entire network information as depicted in Fig. 1(a). The other views differ by the reference user and its neighborhood. Personalized (or local) view follows the omni-guided equal bounds, i.e., limited to two hops in all directions. For example, in personalized view the reference user is represented by black (solid) node. All the nodes that are two hops away from a reference user are considered as known information.

[1] http://www.email-marketing-reports.com/metrics/email-statistics.htm

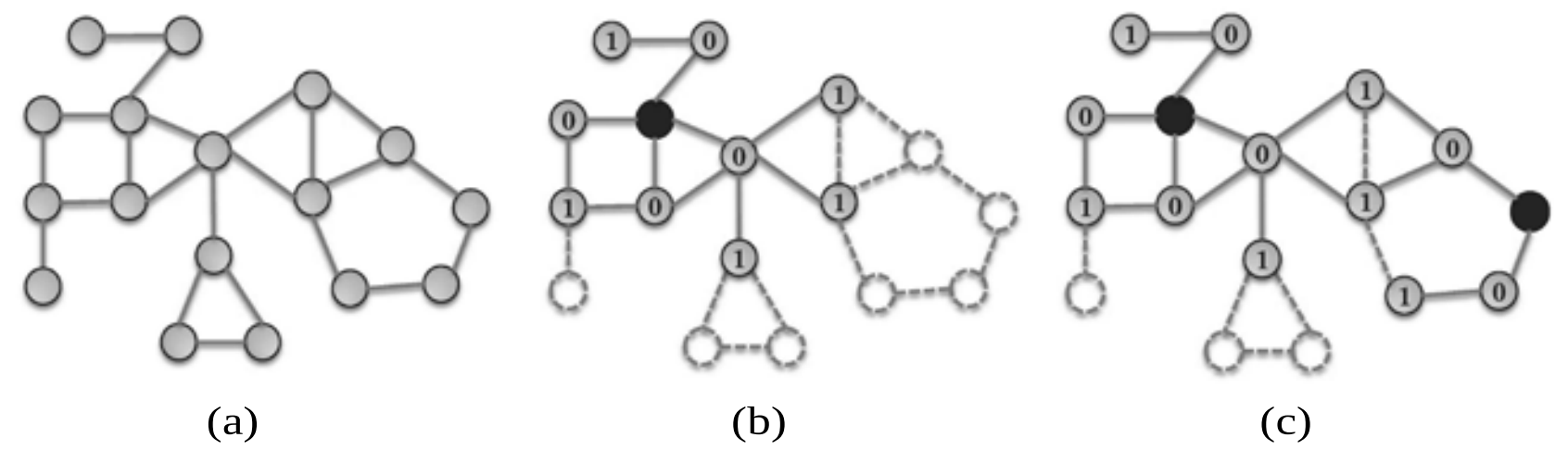

Figure 1: Network Information Perimeter: (a) global, (b) personalized (or local), (c) multi-user personalized

However, communities are identified using extreme views of the email network, from global to local or personalized, which oscillate between efficiency and effectiveness. Moreover, it is almost impossible to approximate the entire email network community structure under the traditional concept of personalization. In order to achieve better communities at reasonable cost we are relaxing the personalized view of the email network using multi-user information. For instance, emails from more than one account can lead us beyond two-hop view of the network as shown in Fig. 1(c). Moreover, there is marginal probability for all the sender/receiver email IDs of each account to be mutually exclusive. It is an interesting and challenging issue to analyze the community structure using multi-user (or multi-account) information under the constraint of privacy. For example, users by exchanging frequent emails through different accounts of the reference user (i.e., owner of the account) with similar behavior are expected to be in the same community. So it requires an effective strategy to explore the multi-account email data for valuable insights in terms of communities.

In this paper, we present a personalized community detection method over multi-user email network, which is solely based on emails extracted from multiple email accounts. Personal emails of each user describe social activities that are transformed to an undirected weighted graph for structural and semantic analysis. Each user, i.e., either sender or receiver, is represented by a node and an edge reflects shared emails, where frequency is associated as an edge weight. The first phase extracts the communication patterns of interest (*CPI*) using multi-user emails as informative features to describe the communication behavior of each user. Subsequently, the second phase detects user communities via an intra-graph clustering method by contemplating structural and semantic aspects together. The semantic resemblance among individuals is achieved through their *CPI*s. We validate the effectiveness of the proposed technique on real email dataset in terms of various performance measures, i.e., density, entropy, and f-score. We also provide comparative analysis on community dynamics in terms of single and multi-user personalized information. Significant contributions of this work, in comparison to existing studies, are summarized as follows:

- **Multi-user Personalized Communities:** We introduce the notion of multi-user personalization under the constraint of privacy and unavailability of an entire email corpus. We uniquely construct an undirected, weighted, and multi-attributed graph using emails meta-data from more than one accounts. This enriched representation of email data enables the generic graph clustering approach to partition the vertices (users) effectively. These user groupings make the email system intelligent enough to filter and group the emails automatically based on similar communication patterns.

- **Community Evolution:** This study uniquely investigates the dynamics of personalized communities in terms of network properties including density, avg. no. of neighbors, network centralization, clustering coefficient, and no. of vertices along with the visual analysis. The community changes is one of the important indicators to detect fraudulent account in email systems.

- **Personalization towards Approximation:** Analyzing the community structure of entire email network of millions of users is computation intensive task for email systems. Multi-user personalization concept provides a mechanism to approximate the entire network community structure with partial information, i.e., a subset of email accounts.

The organization of this paper is as follows: Section 2 emphasizes on the potential applications of personalized communities in the email domain. Section 3 clarifies the concept of multi-user personalization under different circumstances. The details of the proposed method are presented in Section 4. Section 5 contains the empirical analysis of the proposed method on real dataset. Section 6 reviews earlier contributions in relevance to our work. Finally, Section 7 concludes the paper.

## 2. Why Personalized Email Communities Are Important?

Generally, community structures can be easily utilized to facilitate the security mechanism by identifying anomalous behavior and dynamically setting firewall rules (Mcdaniel et al., 2006). However, the subsequent discussion examines the possible applications of discovered communities through personal emails.

### *2.1. Email Strainer*

Due to overwhelming spams attacks (Neumann & Weinstein, 1997), spam filtering has become crucial. Many prior email filtering strategies are limited to widely distributed blacklists, white-listing, and content analysis. Social network dynamics have also been helpful to improve the performance of spam filters (Boykin & Roychowdhury, 2005). Automatic organization of incoming emails, based on some user-specified features, is also a generic approach, where content information, address books and communication patterns are considered to classify emails. Similarly, automatic email prioritization is another well-known type of email filter, which can significantly reduce the amount of effort a user spends manually sorting through emails. Through community structures, the priority level can be easily set for emails based on their presence in a certain grouping. Our proposed method has the ability to fulfill the requirement of a good strainer in the presence of personal (localized) information.

### *2.2. Malicious Email Identifications*

Users with similar communication patterns are automatically grouped, therefore, it is easy to find the related users from the resultant communities. This can aid reactive and preventative measures (Stolfo et al., 2003), and improve the forensic analysis of these events to find the source of infected viruses or groups.

### *2.3. Email Group Predictions*

Community information can also assist in automatic email mailing list generation, which is specific to a particular user. While initiating the communication instance, it can suggest a list of users to be contacted with similar patterns. They could be applied to an automatically generated list of users that may need to be on a given email distribution list. It will also reduce the burden on a system administrator or other list manager.

### *2.4. Guilt by Association*

Identification of communities within other networks, e.g.,telephone or mobile networks, proved highly applicable and beneficial to ascertain fraudulent accounts (Cortes et al., 2001). Examination of the discovered community of a known fraudulent account, with high probability, could determine other fraudulent accounts due to their typical associations. For instance, an organization can use this information to identify accomplices in an unauthorized behavior pattern.

## 3. Multi-User Personalization

The concept of personalization has been studied in literature (Boykin & Roychowdhury, 2004) (Yoo et al., 2009). In order to understand the community structure when dealing with multiple users, i.e., emails from multiple accounts, the traditional personalization concept is ineffective. The main focus of this paper is to consider more than one email account information to group the users based on their analogous communication behaviors.

### *3.1. Personalized Interaction Network (π-Net)*

The personalization term restricts the email interactions to a particular user, i.e., the owner or host of an email account. In other words, we consider all those emails where host user appears in sender or receiver list. We classify the communication among users into active or passive mode, based on who has taken the initiative. If target user is sending the emails to others then it is considered as an active mode of communication. In contrast, when the user is just receiving emails from many other users then it is passive. A typical example of such a personalized interaction network, i.e., $\pi$-Net, is depicted in the Fig. 2.

**Definition 1 (Personalized Emails).** A subset of emails owned by a user from an entire email corpus such that the owner/host appears in the header of each email. The header consists of From, To, CC, and BCC sections.

**Definition 2 (User Role).** Each instance of personalized emails defines the role of a user based on sender and receiver list. Sender becomes active and receivers are considered as passive.

**Definition 3 (User Behavior).** The behavior of a user corresponds to the accumulative nature of personalized email communications to and from that user.

**Definition 4 ($\pi$-Net).** The network constructed from personalized emails such that users are connected with each other based on exchanged emails.

The user roles are tightly coupled with the mode of communication. For example, in Fig. 2(a), a host user A is sending emails to other users, consequently user A is active. However, in Fig. 2(b) the role of user A is passive because no email is initiated by the host. The communication is asynchronous in nature, therefore, user roles are not necessarily mutually exclusive. For instance, in Fig. 2(b) users B, C and D are holding both communication roles. The reason is that at different time intervals these users have different communication roles.

Our intention is to analyze the type of communication among users, rather than the flow of information. We are not intended to study the diffusion process that how to propagate emails effectively in the network. Our goal is to analyze the behavior of users based on their type of interaction and then group them together with similar communication behavior.

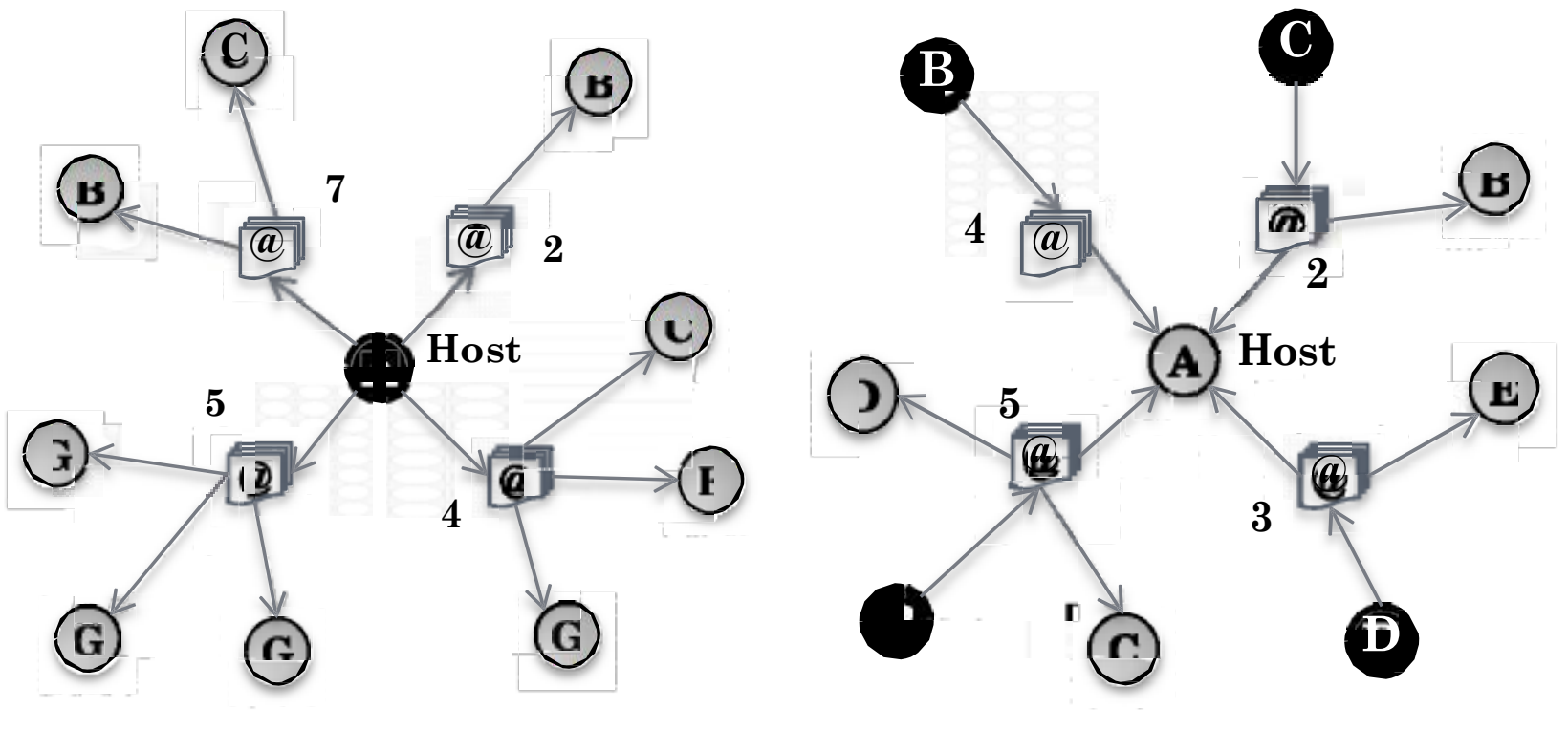

(a) (b)

Figure 2: Personalized Interaction Network ($\pi$-Net) of User 'A' : (a) sender, (b) receiver

### 3.2. $\pi$-Net Varients

The structure of $\pi$-Net varies with respect to nature of the host in the network. It can have more than one host. These $\pi$-Net varients can help us to understand and analyze different aspects of the community structures, e.g.,group of users to whom the host has similar interaction through various social roles like relative or workmate.

The semantics of multi-user term in this paper is beyond trivial definition. It has dual meaning based on the scope of email information, i.e., network information perimeter given in Fig. 1. First, a user can have multiple email accounts for different purposes irrespective of the service providers. Second, emails from multiple accounts where each user has one exclusive email account. Therefore, the boundary of email information is limited to the emails from (i) multiple accounts of a single user (ii) multiple users having their own single account. In both cases, single strategy can effectively produce groupings of users with varying semantics.

It is also interesting to analyze the community structure in diverse environment where network is constructed from multiple email accounts of a single user as a single host i.e., integrated or fused $\pi$-Net. It is the special case of multi-user $\pi$-Net. The diversity comes from heterogeneous nature of user's interactions, e.g., user can exchange emails as a friend, relative, or co-worker. Subsequent discussion demonstrates the approach to construct semantic rich $\pi$-Net and then identify communities.

## 4. Multi-User Personalized Email Communities

The ultimate goal of discovering multi-user personalized communities is to portray the grouping of individuals with similar neighborhood and communication behavior. The abstract idea is already presented in (Nawaz et al., 2013).

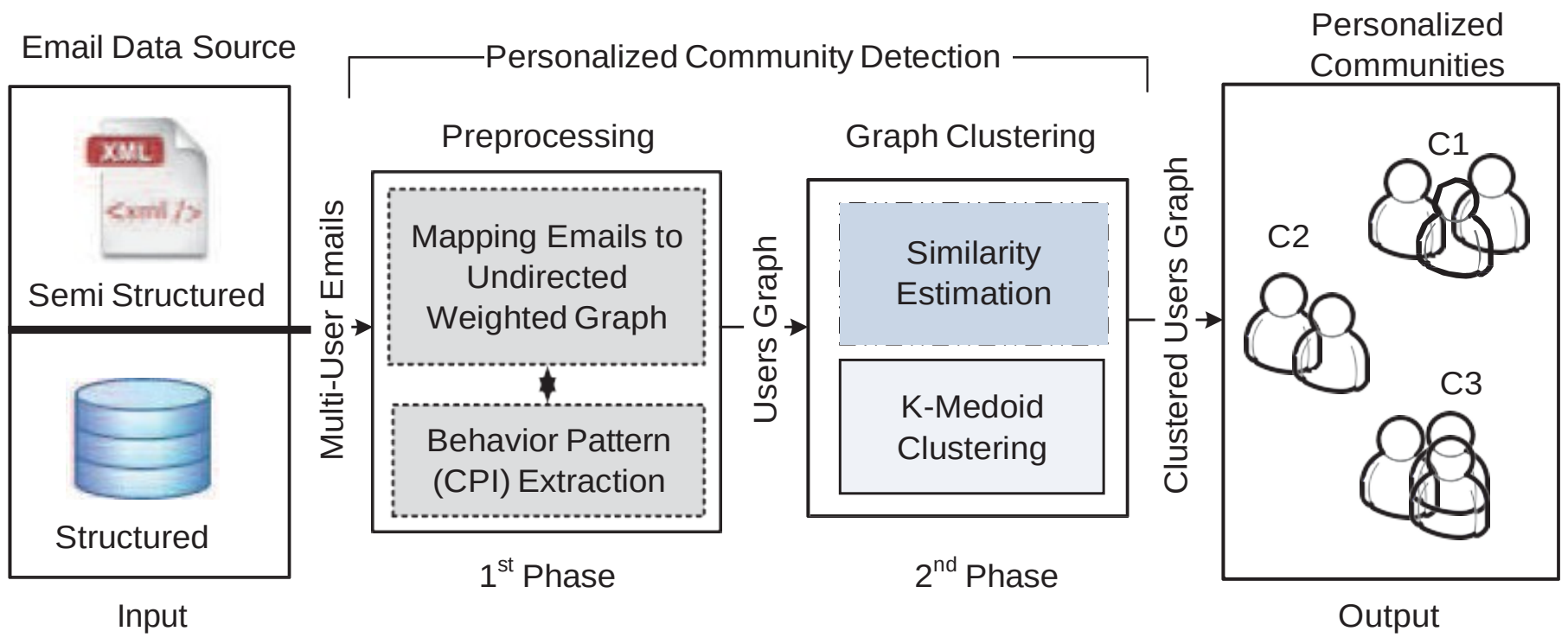

Figure 3: System Architecture for Personalized Community Detection from Multi-User Emails.

The architecture of the proposed method with two mutually exclusive, sequentially dependent phases is given in Fig. 3.

Initially, multi-user information is acquired from more than one email accounts that consists of electronic communication among people, i.e., multi-user $\pi$-Net. In the preprocessing phase, the interactions among the users are represented in the form of an undirected weighted graph (UW-Graph). A set of attributes are attached with each node to reflect the user's communication pattern or behavior. In the second phase, the groups of individuals are identified based on their overlapping communication interests using an intra-graph clustering approach (Nawaz et al., 2012), which is simple and effective in this context. It has the capability to capture both topological and semantic aspects concurrently during the clustering process.

### *4.1. Preprocessing Phase*

At the beginning, emails from multiple user accounts are provided as an input to the system. This information is directly extracted from personal emails of corresponding user respository under the constraint of privacy. Moreover, it is represented as a UW-Graph describing behavioral patterns for further analysis, unlike the case with prior techniques. Even though personalization can lead us toward an approximation of an individual's behavioral activities, our focus is to attain a global perspective within a confined knowledge domain.

#### *4.1.1. Transition from Multi-user Emails to Social Graph*

An email network has a graph structure whose nodes or entities represent either individuals or organizations, and edges represent their email communications. It is trivial to map the data from a real email network onto a UW-Graph. However, it is essential to use robust graph mining algorithms. Our intention is to represent this email network as a graph for better community analysis. The graphical description of the transition or mapping strategy from personalized directed email network,i.e., $\pi$-Net, to a UW-Graph is given in Fig. 4. An email network consists of users and shared emails. An arrow indicates the flow of information among individuals in the network. All the emails shared between an arbitrary pair of users can be associated with directed links.

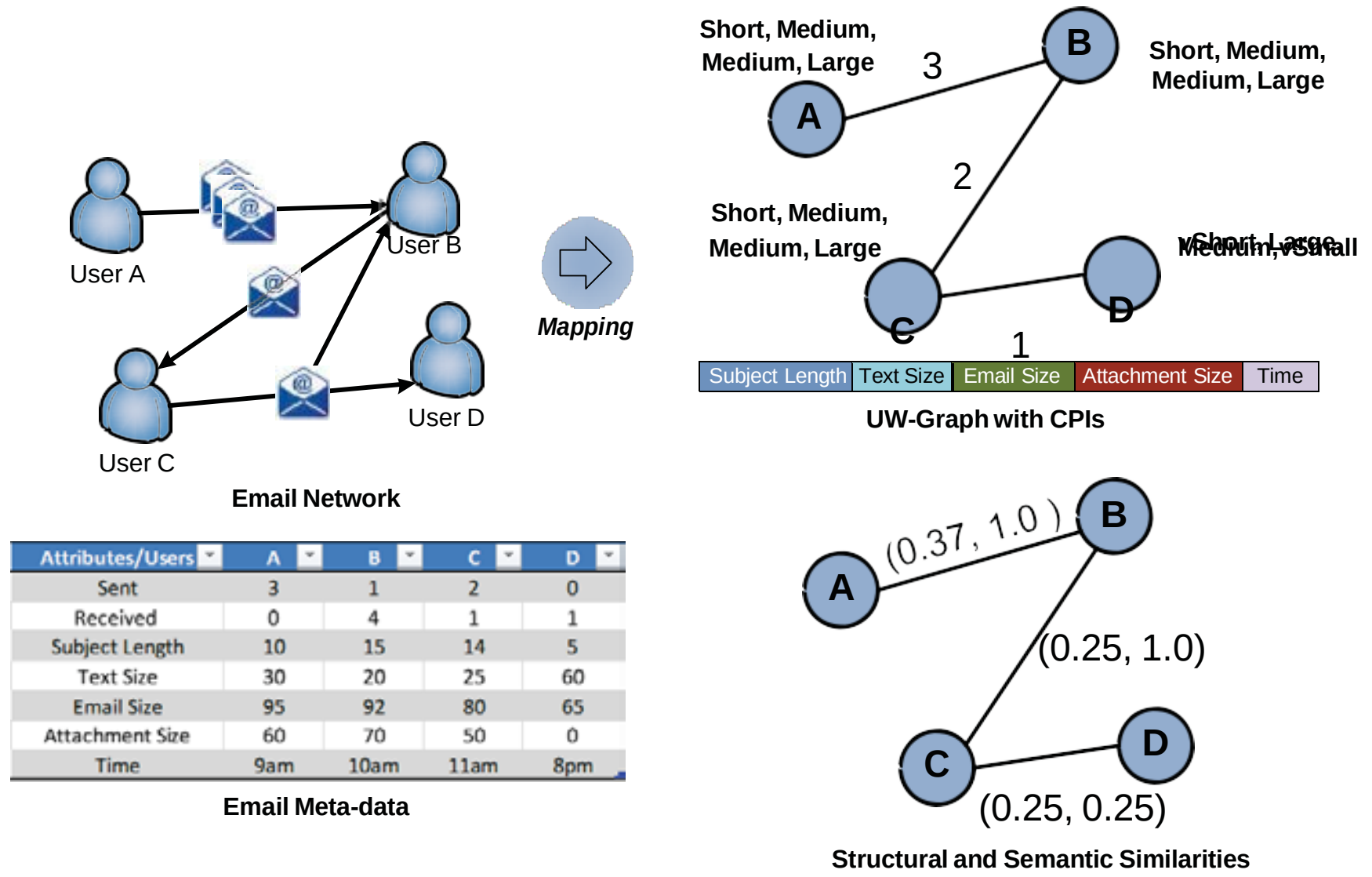

| Attributes/Users | A | B | C | D |
|---|---|---|---|---|
| Sent | 3 | 1 | 2 | 0 |
| Received | 0 | 4 | 1 | 1 |
| Subject Length | 10 | 15 | 14 | 5 |
| Text Size | 30 | 20 | 25 | 60 |
| Email Size | 95 | 92 | 80 | 65 |
| Attachment Size | 60 | 70 | 50 | 0 |
| Time | 9am | 10am | 11am | 8pm |

Figure 4: Multi-user Personalized Emails towards UW-Graph.

In the graph domain, each user is represented by one vertex and the intensity of emails exchanged is reflected by link weights. The knowledge extracted from a directed email network has different semantics compared with an undirected email network. We do not process the real email network to analyze the flow of information in the network. Our focus is to capture an individual's behavior with regard to information exchange. We construct personalized email network from outgoing emails, as suggested by Steve Martin et al. (Martin et al., 2005) in relevance to behavior analysis. The emails are either sent to/from reference user, who owns the email account. For example, in Fig. 4, User B owns the email account. In this case, if user A exchanges any email with user C or D then it will not be considered in this network due to privacy constraint. The topological structure of a personal real email network is visualized [2] shown in Fig. 5 along with the designation mapping of each user. It is possible to have multiple accounts for a particular user, e.g., Sally Beck. The size of each vertex represents the volume of emails exchanged with other users.

[2] An open source software platform for visualizing complex networks: http://www.cytoscape.org/

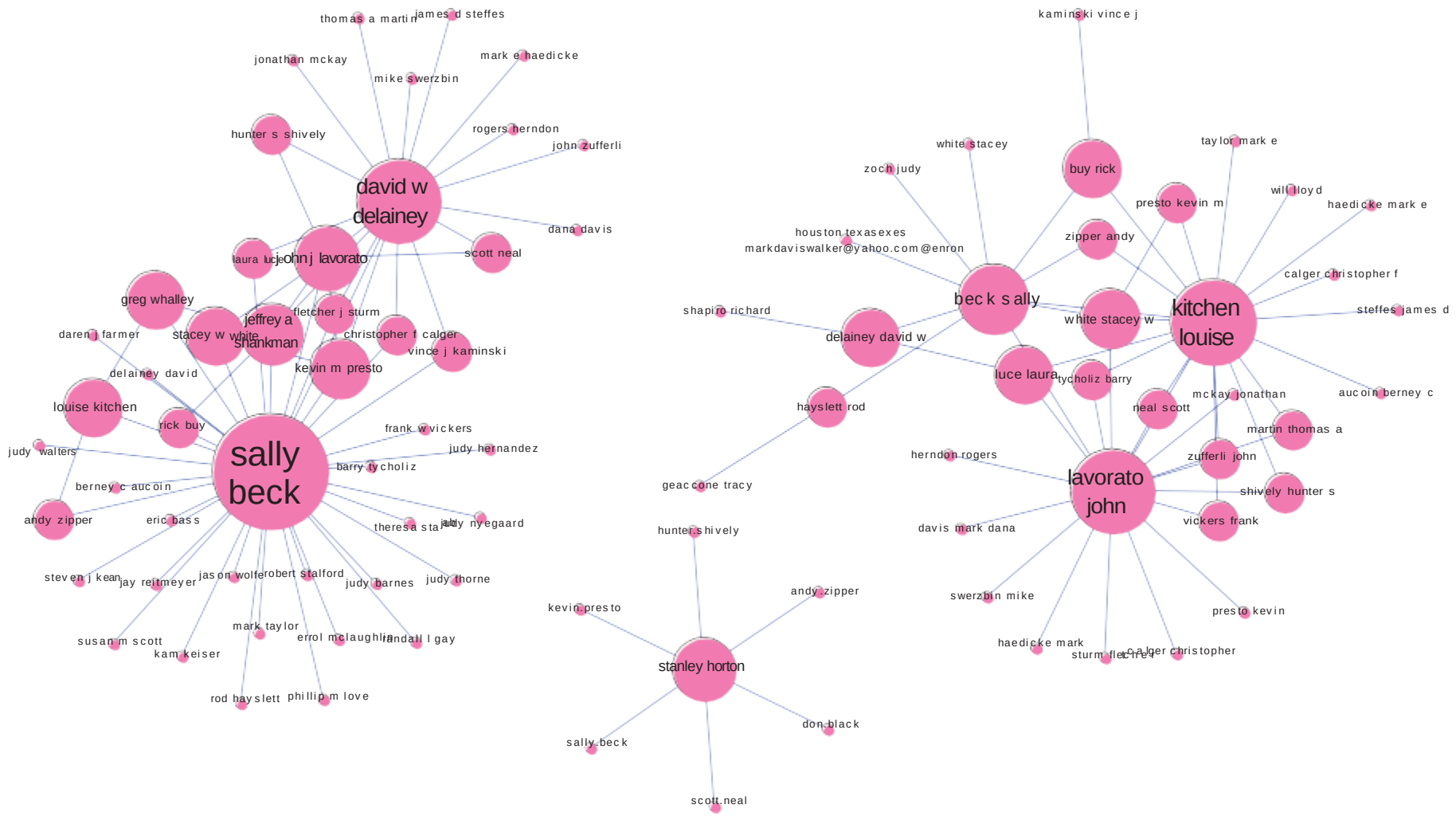

(a)

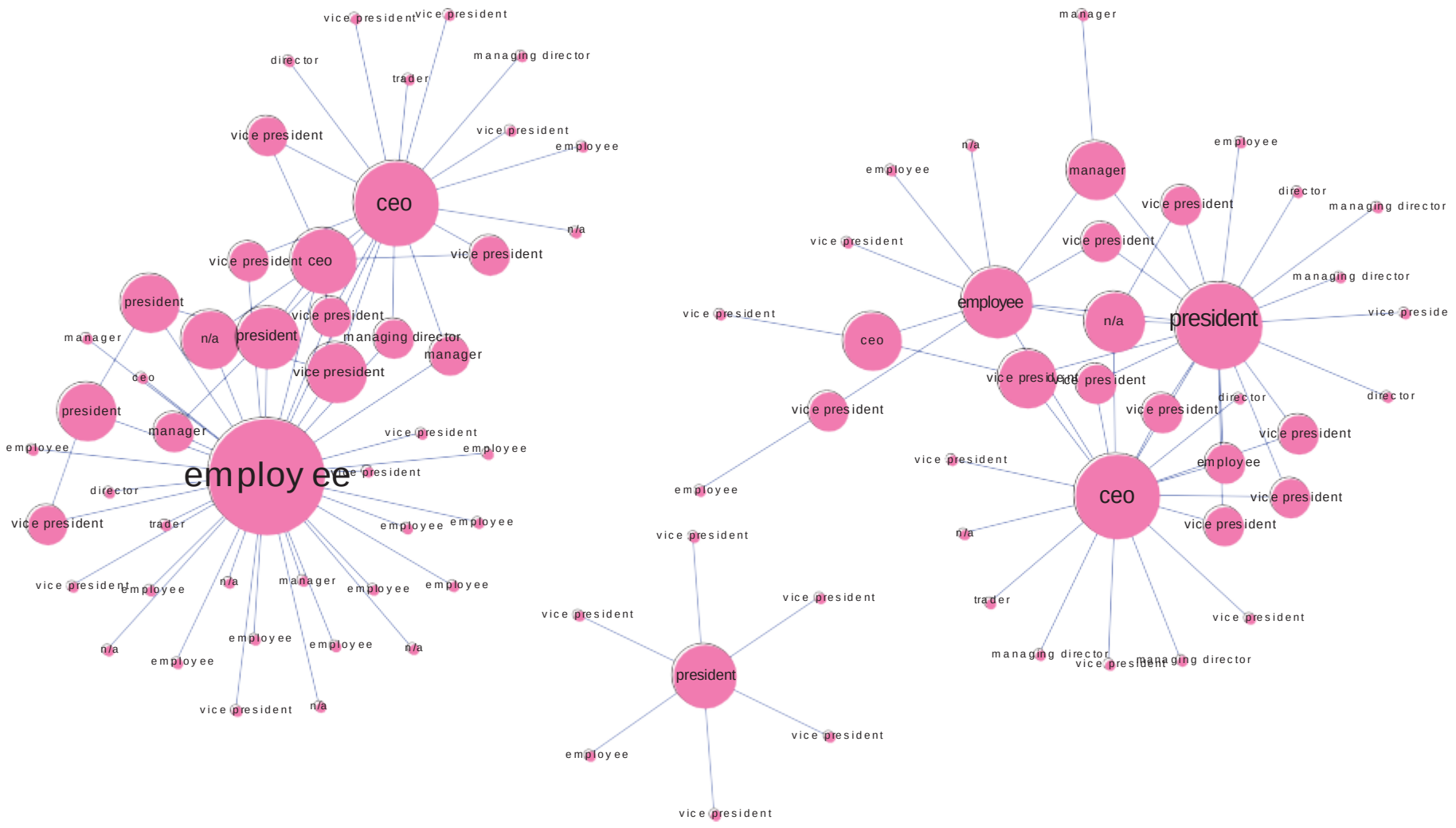

(b)

Figure 5: Sally Beck's Multi-account Email Network (a) Original interaction (b) Mapping user names to their designations

An email contains the sender and receiver information along with the content's meta-data which is, in contrast to traditional approaches, an essential entity for us to determine the level of communication interest among different individuals. The meta-data from an email's content includes subject length, text size, and attachment size. This meta-data forms the semantics of an email, as depicted in Fig. 4. The topological transition process is trivial compared with anticipating the representative semantics. The subsequent discussion highlights the technical aspects of the graph along with some challenging issues.

#### *4.1.2. CPI Extraction: Communication Behavior Patterns*

The behavioral aspect uniquely depends on email's meta-data(Martin et al., 2005), i.e., a set of edifying features which is associated with each vertex. The collection of selective properties or meta-data of an email constitutes one complete communication pattern of interest (*CPI*), as given in Eq. (1). A distinct email communication between two users is represented by a vector (or pattern/*CPI*) where each element corresponds to a particular email property or attribute.

Generally, individuals exchange many emails among themselves. The diversity of users and shared information through one or more emails are the ultimate challenges. To cope with the diverse nature of emails, we have investigated the cumulative effect by considering the most prominent *CPI*. More precisely, the influential instance of each element of the *CPI* is determined to construct the representative *CPI*, i.e., *Influential CPI*, for an arbitrary user. Consequently, the behavioral aspect of a particular user can be determined through an *Influential CPI*, which is evident from Eq.(2).

$$CPI = \{a_i | i = 1, \ldots, N_a\} \tag{1}$$

$$\text{Influential } CPI = argmax_j\{Frequency(a_{ij})\} \quad \text{where } j = 1, \ldots, n_{ai}. \tag{2}$$

The value of $i^{th}$ attribute $a_i$ is expressed as $a_{ij}$, $N_a$ refers to the total number of attributes, and $n_{ai}$ represents possible values for each attribute $a_i$. *Influential CPI* is based on the most frequently occurring attribute values in all of the communication from a particular user. In case of collision the selection is based on the frequency of other attributes. The frequency concept overwhelms a user's diversity. On the other hand, a distinct set of email features have given us a chance to reflect the variety of emails via *Influential CPI*. Moreover, the heterogeneous attributes related to email require further consideration.

All the properties of an email are good candidates for constructing a *CPI*. However, it is important to investigate the properties significantly contributing to form user communities with homogeneous behavior. Fortunately, in this domain, Steve Martin et al. (Martin et al., 2005) have already analyzed the influence of mutually exclusive email properties. Accordingly, significant email properties are used to construct an *Influential CPI*, as briefly explained below:

- **Subject Length:** Number of characters in the email's subject provides assistance to capture the basic user profile to imitate the writing in the email.

- **Text Size:** All the text written or contained in an email is also a distinctive feature. When an email gets forwarded repeatedly, the size of the body text gets inflated. This can help us to determine which individuals are frequently receiving or sending forwarded emails. Consequently, the type of information shared within the network can be easily detected, which allows us to readily identify influential individuals.

- **Attachment Size:** All the documents or files attached with the email can help us to find the key users in the network. Announcements or messages sharing activities are less helpful. Those who are involved in abnormal activities or sharing large amounts of data through emails can also be identified. Therefore, it is limited to specific scenario.

- **Email Size:** It reflects the overall size of the email including attachments. It has potential to depict the nature of content that is shared among users. Sometimes, users share non-textual content inside the email body that can not be easily identified with other properties.

- **Date and Time:** An email is given a time/date stamp when users exchange it with others. It is the most discriminative feature in this context. Individuals have email communications in a certain time period. It is ubiquitous in nature and directly affected by the personal life of a user. Consequently, it has a vital role in identifying users with similar behavioral patterns with respect to their communication time-line.

Aside from the topological structure inferred through interaction, conceptually all of the above email properties can make significant contributions to the overall community detection process. We are trying to bypass in-depth analysis of email content, which requires computation intensive natural language processing. Yet, there are a few challenges to construct the $CPI$ and require our attention. Most importantly, all the listed properties are numeric except for the time-stamp. Their mutual analysis demands homogeneity in interpretation, which can be achieved through scaling.

All the properties get transformed from real (actual values) to nominal (descriptive) or ID spaces to ensure proper description and efficient processing. Accordingly, the time stamp of an email is initially divided into mutually exclusive communication time slices (CTS) and then assigned labels based on daily life greetings[3]. These labels are further mapped to IDs, as shown in Fig. 6. The parallel coordinate concept is adopted to depict this process, which is reversible in nature. The granularity level throughout in this paper is selected on a daily basis for CTS. The remaining attributes get scaled to nominal and id spaces in a similar fashion. The nominal space for each attribute is determined intuitively. In each space, the minimum and maximum values are considered to determine the boundaries while the number of intervals remains constant for brevity.

[3] https://answers.yahoo.com/question/index?qid=20090529125255AA2Xvcd

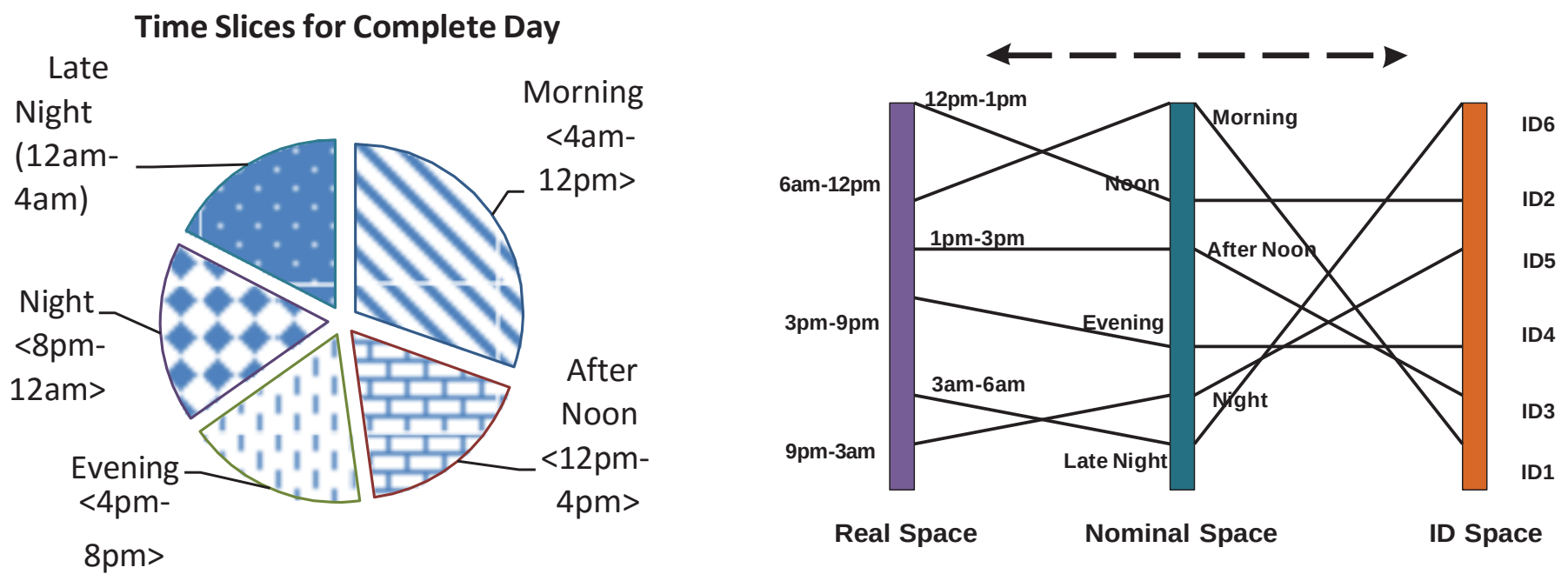

Figure 6: CTS and Mapping across Real and ID spaces.

$$TAG = \{T_a | a \in \{SubLen, \; TxtSize, \; EmailSize, AttachSize, \; Time\}\} \quad \text{where } T_a = \{tag_i | i = 1, \ldots, n_a\} \tag{3}$$

Each attribute has a set of labels,$T_a$ that are linked with the corresponding value in the other space. $tag_i$ is the actual tag or label. The total number of labels, $n_a$, depends upon the versatility of the property. On the other hand, the set of attributes and their labels, i.e., $TAG$, remains fixed as given in Eq. (3). Finally each $CPI$ contributes to form the *Influential CPI* of a particular person in the network. Once we get the topological structure and associated $CPIs$, then a kind of social graph is formed. From now on, we can employ a graph mining algorithm to this social graph/network for discovering social communities.

### 4.2. Community Detection through Graph Clustering

In order to discover underlying user community structures from this social graph, we use an existing graph partitioning approach, i.e., intra graph clustering (Nawaz et al., 2012). It has the capability to group all the multi-attributed nodes with diverse neighborhood structure. In other words, this partitioning strategy considers both the topological structure and semantic resemblance, through $CPI$, among nodes. Ultimately, we obtain groups of email users based on their communication interactions and behavior.

An undirected, weighted, multi-labeled graph $G = \{V, E, W, A\}$, not necessarily connected, where $V = \{v_1, v_2, ...v_N\}$, $E$, $W$ and $A$ are the set of vertices, edges, weights, and attributes respectively. An arbitrary pair of vertices $v_i$ and $v_j$ are connected through an edge $e_{ij} \in E$ with each other having non-negative cost $w_{ij} \in W$. Each vertex $v \in V$ contains $M$ attributes, i.e., $A_v = \{a_v^1, a_v^2, ...a_v^M\}$. $a_v^i$ represents the value of $i^{th}$ attribute of a vertex $v$. The total number of edges incident on a vertex $v_i$ are represented as its degree $deg_i$. Two vertices $v_i$ and $v_j$ are direct neighbors if there exists an edge between them, i.e $e_{ij} \in E$ , otherwise, we consider them as indirect neighbors.

**Definition 5 (Path).** A path $p_{ij}$ between an arbitrary pair of vertices $v_i$ and $v_j$ is a sequence of vertices $p_{ij} = \{v_i = v_1, v_2, \dots, v_l = v_j\}$, such that $e_{ii+1} \in E$ and $i \in [1, l-1]$.

**Definition 6 (Shortest Path).** The shortest path $sp_{ij}$ between an arbitrary pair of vertices $v_i$ and $v_j$ is a path with least summation of edges cost.

Intra graph clustering is the process of relative partitioning of the vertices of the graph into $K$ disjoint sub-graphs where $G_i = \{V_i, E_i, W_i, A_i\}, V = \cup_i^k(V_i)$ and for any $i \neq j$, $V_i \cap V_j = \varphi$. Initially, at the first step, we estimate the rela- tionship among nodes in an un-clustered graph. The relevance among nodes is calculated in a collaborated way using topological and contextual information. The second step groups (or cluster) the similar vertices using k-medoid frame- work. The topological structure and associated contextual information of the graph remains unchanged during clustering process. Therefore, similarity val- ues among vertices are calculated once and utilized multiple times in successive steps.

### 4.2.1. User Relevance Estimation using Collaborative Similarity Measure

The similarity measure determines the strength of relationship or connectiv- ity among pair of vertices or users. The proposed method utilizes this strength to group or cluster similar vertices together. We define the similarity measure for weighted, undirected, and multi-labeled graph; where the connectivity rela- tionship among pair of vertices has the following forms: (1) Directly Connected, (2) In-directly Connected, (3) Disconnected. The directly connected pair of ver- tices share an edge between them. When there is no direct link between two vertices, e.g., $v_i$ and $v_j$, we can reach vertex $v_j$ from vertex $v_i$ by following an arbitrary path. In the absence of direct and indirect links, vertices are stated as disconnected. In this paper, we assume that two directly connected vertices share a single link.

$$SIM(v_a, v_b)_{struct} = \begin{cases} \dfrac{w_{ab}}{\sum_{c=1}^{deg(v_a)} w_{ac} + \sum_{c=1}^{deg(v_b)} w_{bc} - w_{ab}}, & directly\ connected \\ \prod_{i=snode}^{dnode} SIM(v_{pi}, v_{pi+1})_{struct}, & in-directly\ connected \\ 0, & disconnected \end{cases} \quad (4)$$

The collaborative similarity among an arbitrary pair of vertices, source to destination in a graph, is computed through structural and contextual simi- larity inspired by Jaccard similarity coefficient (Jaccard, 1901). In Eq. (4), $SIM(v_a, v_b)_{struct}$ represents the structural similarity between two vertices, $v_a$ and $v_b$, by exploiting the neighborhood of both vertices and strength of their direct interaction. In other words, it is defined as the weighted ratio of common neighbors to all the neighbors of both vertices. The overall structural similarity is proportional to the number of common neighbors. A large number of common neighbors yields high pair-wise similarity value. A directly connected pair of vertices employ immediate neighborhood information to estimate the similarity value. However, the similarity value for indirectly connected vertices, by following a path, is calculated through linear product of direct structural similarity values. The similarity value becomes zero for disconnected vertices. This strategy is applicable on weighted and un-weighted graphs. In the absence of edge weights, constant value 1 is expected to be associated with each link in the graph.

One of the key aspects of this approach is to consider contextual similarity to attain structural cohesiveness among nodes that is defined in Eq. (5) and Eq. (6). Its importance is evident from the applications where the nodes emerge in different contexts. For instance, in social network, the users are represented by nodes and edges reflect their relationships. Each user can have different roles or contexts like occupation as student, doctor, engineer, or designer. We consider fixed number of attributes for each user to define the context.

$$SIM(v_a, v_b)_{context} = \begin{cases} \frac{\sum_{i=1}^{M} COMMON(v_a, v_b, a^i) * (w_{a^i})}{\sum_{j=1}^{M} w_{a^j}}, & otherwise \\ \prod_{i=snode}^{dnode} SIM(v_{pi}, v_{pi+1})_{context}, & indirectly\ connected \end{cases} \quad (5)$$

$$COMMON(v_a, v_b, a^i) = \begin{cases} 1, & \textbf{if } v_a \text{ and } v_b \text{ hold same value on } i^{th} \text{ attribute} \\ 0, & otherwise \end{cases} \quad (6)$$

$$CSIM(v_a, v_b) = \alpha * SIM(v_a, v_b)_{struct} + (1 - \alpha) * SIM(v_a, v_b)_{context} \quad (7)$$

We also support prioritization of vertex semantics by associating attribute weight $w_{a^i}$ in Eq. (5) and Eq. (6). At beginning of the algorithm, these values get initialized and remain fixed throughout all computations. The combine effect of structure and context relevance among vertices is defined in Eq. (7). The parameter $\alpha$ is introduced to control the influence of both similarity aspects. The importance of structure and context similarity is variant and depends on the application domain. Therefore, the choice of appropriate value for this parameter is critical. For instance, social networks often exhibit dense regions and follow power-law degree distribution. The higher value for $\alpha$ seems effective for these networks because nodes in dense regions are expected to have similar attributes. However, non-scale free networks, e.g., road networks, need to be treated with a balanced ratio.

The collaborative similarity value is calculated for indirectly connected vertices by following a path. However, a pair of vertices can have multiple paths. We choose the shortest path as a candidate path to estimate the similarity value. In order to extend the similarity using shortest path approach, we must utilize the desirable property, i.e., similarity (distance) value decreases (increases) as we move far from the source vertex. We achieve this by taking the reciprocal of similarity measure by defining a distance function in Eq. 8. The distance value in close proximity is expected to be low due to transitivity property. In a weighted graph, shortest path between two vertices may not be unique. We pick the one with least distance value at initial expansions.

$$DIST(v_a, v_b) = \begin{cases} \frac{1}{CSIM(v_a,v_b)}, & \text{directly connected} \\ \sum_{i=snode}^{dnode} \frac{1}{CSIM(v_{pi},v_{pi+1})}, & \text{in-directly connected} \\ \infty, & \text{disconnected} \end{cases} \quad (8)$$

#### 4.2.2. Grouping Users based on K-Medoid Framework

The users are grouped together using a K-Medoid clustering framework. Systematically, it consists of two main components: similarity calculation and clustering, which is iterative in nature. This module requires two parameters other than the social graph: the number of clusters to be found and an influence factor which controls the importance of the topological structure and semantic associations. It does not require altering the structure of the graph. A brief description of the clustering module is given in Algorithm 1.

An initialization step considers the issues related to memory allocation as well as variable management for temporary and permanent storage. Social graph data is retrieved and stored in the main memory for later processing. Similarity value estimation is a core part of this algorithm due to dependence of subsequent processing. Basically, the similarity value is anticipated among all possible pairs of vertices in terms of their topological and behavioral or semantic resemblance, using Eq. (7).

At this point, the most important factor that needs to be considered is the symmetry of the values due to the undirected graph. For example, if there are two vertices $v_a$ and $v_b$ then DIST($v_a$,$v_b$) = DIST($v_b$,$v_a$). At the last step, the partitioning of the graph vertices is done by utilizing the pre-calculated similarity values among vertices by employing the inherent features of the K-Medoid clustering infrastructure. To start, vertices are randomly selected as the centroids (expected center points) or initial seeds to represent the hidden clusters. Then, we associate neighboring vertices to the nearest centroids to create a partition based on their similarity distance. Finally, the quality of each cluster is analyzed based on density and entropy.

**Algorithm 1** Social Graph Clustering

**Require:** A social graph $G = \{V, E, W, A\}$, no. of clusters $K$, weight factor $\alpha$
**Ensure:** $K$ clusters $V_1...V_i...V_K$, where $V_i$ contains all the vertices in cluster $i$.

1. **Initialization**, $DIST[i,j] = 0, iteration = 0$, $cList[] = 0$, $w_{a^1}, \dots w_{a^M} = 1$, where $(i, j) = 1 \dots N$
2. **Distance Calculation**,
   (a) **for** each vertex pair $v_i$,$v_j$ in $V$ where $(i,j) = 1 \dots N$ and $j \neq i$ **do**
   $DIST[i, j]$ = Calclute distance using Eq. 8.
   (b) **end for**
3. **Social Community Detection through K-Medoid Clustering framework**, {Initially choose centroids with high degree}
   (a) Top $K$ vertices as initial centroids for each cluster with respect to degree $cList = TopK(V)$
   (b) **for** each vertex $v_i$ in $V$ **do**
   $cluster[i] = argmin_{i,j}\{DIST(i, j)\}$ for all centroids $j = 1 \dots k$
   (c) **end for**
   (d) Evaluate the clusters quality through density(Eq. 9) & entropy (Eq. 10)
   (e) **if** Converge || Iterations are maximized **then**
   return $K$ clusters $V_1 \dots V_i \dots V_k$
   (f) **end if**
   (g) Update $centroidList[j]$ for each cluster $j = 1 \dots k$, for which the sum of distances is minimum, using Eq. 8.
   (h) Repeat steps 3$b$-3$g$ untill convergence

## 5. Experiment

In this section, user communities are identified using the proposed method in the context of a personalized email dataset. Usually, it is very hard to justify the clustering results, i.e., communities, due to the absence of ground-truth. However, these communities are validated against three performance measures. The proposed method has been implemented in Java[4].

### 5.1. Email Dataset

In real life, due to email's privacy issues, there is no public corpus from a real organization available except for a huge Anonymized Enron email corpus (Shetty & Adibi, 2004). It contains vast collection of emails covering a time span of 41 months, and also uniquely depicts the ups and downs of the energy giant Enron. It provides an opportunity to determine related mailbox users based on their unique communication and relationship in the email network. We have considered the Enron email data-set[5], which contains all the emails of 161 users, managed separately, to infer the community structure from partial information available in terms of personalized emails. Two user's, Sally Beck and Louise Kitchen, email network is exploited from this data-set for all the experiments in this paper to infer the community structure. The email interactions with the individuals outside the Enron Corporation are explicitly ignored to reflect factual associations.

### 5.2. Community Validation Methods

The multi-user personalization concept does not require an explicit instance of validation measure for communities with respect to connectivity and semantics. Therefore, state of the art community validation measures are used to analyze the competitive nature of the proposed algorithm. The structural and contextual quality of the communities has been analyzed in terms of Density and Entropy measures (Cheng et al., 2011) respectively. However, the combined impact local and global semantics is also examined through F-Measure (Witsenburg & Blockeel, 2011).

#### 5.2.1. Density: Topological Structure

A strong connection among users can be easily analyzed by utilizing the density function. It is the ratio between the number of links present in a community to links contained in all communities. The ratios accumulate for all the communities to evaluate the overall impact. Its values lie in the interval of [0, 1].

$$Density = D\left(\{V_q\}_{q=1}^{k}\right) = \sum_{q=1}^{k} \frac{|\{(v_a, v_b) \mid v_a, v_b \in V_q, (v_a, v_b) \in E\}|}{|E|} \quad (9)$$

[4]Proposed method's binaries will be provided on request (http://sites.google.com/a/dke.khu.ac.kr/wnawaz)

[5]http://www.edrm.net/resources/data-sets/edrm-enron-email-data-set-v2

Where $q = \{1, 2, \ldots k\}$ clusters.

#### 5.2.2. Entropy: Semantics

One of the key aspects to measure the quality of the cluster is to determine the relevancy among vertices based upon their attributed nature. For each attribute the small entropy, in Eq. (10) , is calculated against each cluster with associated attribute and product with weighted ratio of the attribute.
The percentage of vertices in the cluster having respective $n^{th}$ value on $i^{th}$ attribute in cluster $q$ which is represented by $Prcnt_{iqn}$. When all the vertices inside the same cluster are having similar attributes or contexts associated with them, then overall entropy acquires minimum value.

$$Entropy = E\left(\{V_q\}_{q=1}^{k}\right) = \sum_{i=1}^{M} \frac{w_{attr_i}}{\sum_{j}^{M} w_{attr_j}} \cdot \sum_{q=1}^{k} entropy(attr_i, V_q) * \frac{|V_q|}{|V|} \tag{10}$$

$$entropy(attr_i, V_q) = -\sum_{n=1}^{|Dom(attr_i)|} Prcnt_{iqn} log_2 Prcnt_{iqn}$$

Where $q = \{1, 2, \ldots k\}$ clusters, $n = \{1, 2, \ldots |Dom(attr_i)|\}$ attribute values.

#### 5.2.3. F-Measure: Local & Global Semantics

The f-measure has the ability to evaluate the collective (i.e., local and global view) qualitative nature of the formed cluster. Precision and recall are the two dominating factors in this measure. Precision can be defined as the ratio between number of vertices have attribute $a$ for cluster $q$ and total number of vertices in that cluster.

$$FMeasure = \frac{1}{M} \sum_{i=1}^{M} \left( max_q \left( FM_q^{attr_i} \right) * v_q^{attr_i} \right), q = 1 \ldots k \tag{11}$$

where

$$FM_q^{attr_i} = \left(2 * P_q^{attr_i} \cdot R_q^{attr_i}\right) / \left(P_q^{attr_i} + R_q^{attr_i}\right)$$

and

$$Precision = P_q^{attr_i} = \frac{v_q^{attr_i}}{|V_q|}$$

$$Recall = R_q^{attr_i} = \frac{v_q^{attr_i}}{|v^{attr_i}|}$$

Recall is estimated in the same way. The only difference is the denominator, where we consider all the users in the entire graph with associated attributes rather than considering the boundary of the community. Precision is relatively local compared to recall, which has global impact. For all attributes introduced in the results, the maximum value for precision and recall ratio is considered in Eq. (11). The possible value for this measure is between 0 and 1, where 1 is the best score.

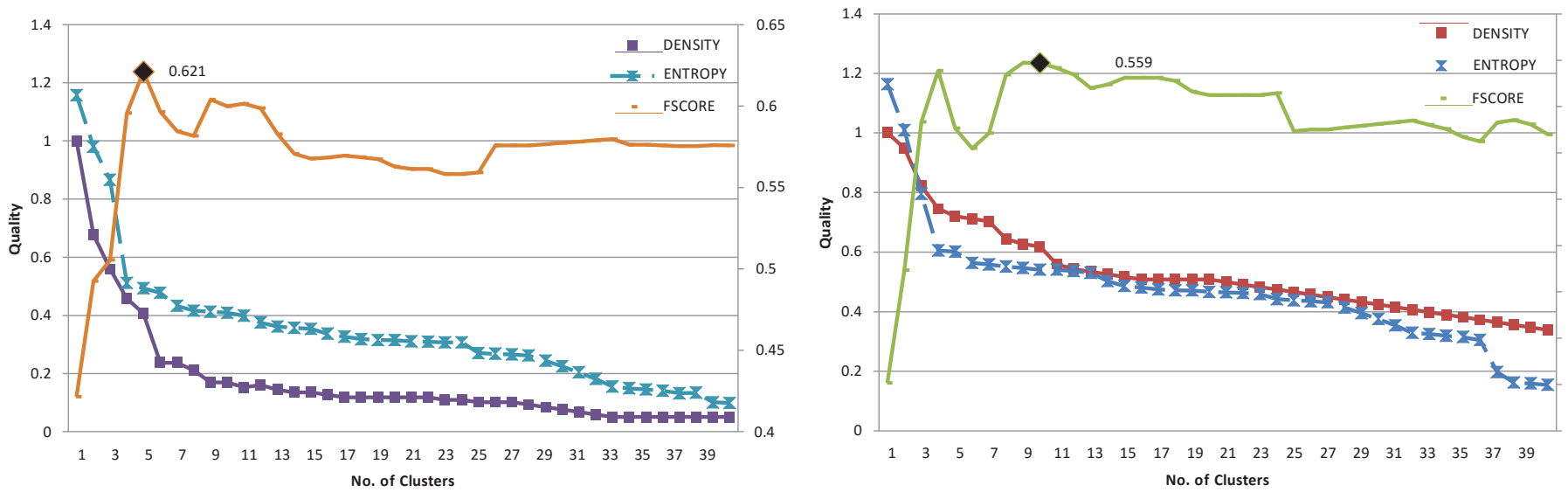

(a) (b)

Figure 7: No. of Communities vs. Quality: Sally-Beck's Personalized Communities [(a)multi-account, (b)single-account/fused $\pi$-Net], $\alpha = 0.5$

### 5.3. Result Analysis

We study the strength and effectiveness of the multi-user personalized communities under the constraint of quality, community dynamics, and network exploration. Initial discussion highlights the qualitative analysis of the results using state-of-the-art measures, as explicated in the former section. Later on we analyze the dynamics of email network and its exploration using single and multi-user personalized information along with visual description.

#### 5.3.1. Qualitative Analysis

The members of a community should reflect strong behavioral bonding with each other. We analyze the strength and nature, i.e., density and entropy respectively, of the emails exchanged among individuals as a measure to identify behavioral bonding. Intuitively, it has strong relationship with the number of communities to be identified. Therefore, we plot the qualitative values with respect to varying number of communities in Fig. 7 to understand this phenomenon. It is not surprising to see the linear decline of result's quality with the number of communities. However, beside the fact that density and entropy are inversely proportional to each other, it is interesting to observe the contrast among density and its counterpart, i.e., entropy and f-measure. For instance, in Fig. 7(b), the global maxima for f-measure reflects relatively strongly communication among individuals with similar behavior when number of communities are low, i.e., $K = 5$. The optimal value for $K$ is somehow subjective to the nature of the network. It is evident from Fig. 7(a) that Sally Beck's multiple accounts are considered independently. Fig. 8 visualizes evolution of communities with multi-account information.

Prior to clustering phase, we estimate similarity among two arbitrary users by accumulating their weighted topology and semantics through a controlling parameter $\alpha$. We can easily control the contribution of each aspect of the overall similarity estimation by adjusting the Alpha value. For example, if we intend to consider only the structural aspect while neglecting the semantics, then the value should be equal to 1. The alpha value, i.e., either 0 or 1, reflects the importance of each aspect in similarity estimation. However, its value between 0 and 1 considers the partial contribution of each aspect. Moreover, we analyze this parameter against the f-measure of each resultant community by varying the number of communities in Fig. 9. It is evident that an Alpha value of 0.5 can produce better quality results. Considering this fact, all the results in Fig. 7 have equal weightage on both aspects, i.e., $\alpha$ is 0.5. Further, we can anticipate the optimal value of $K$ by varying its value at clustering phase to identify the best quality results.

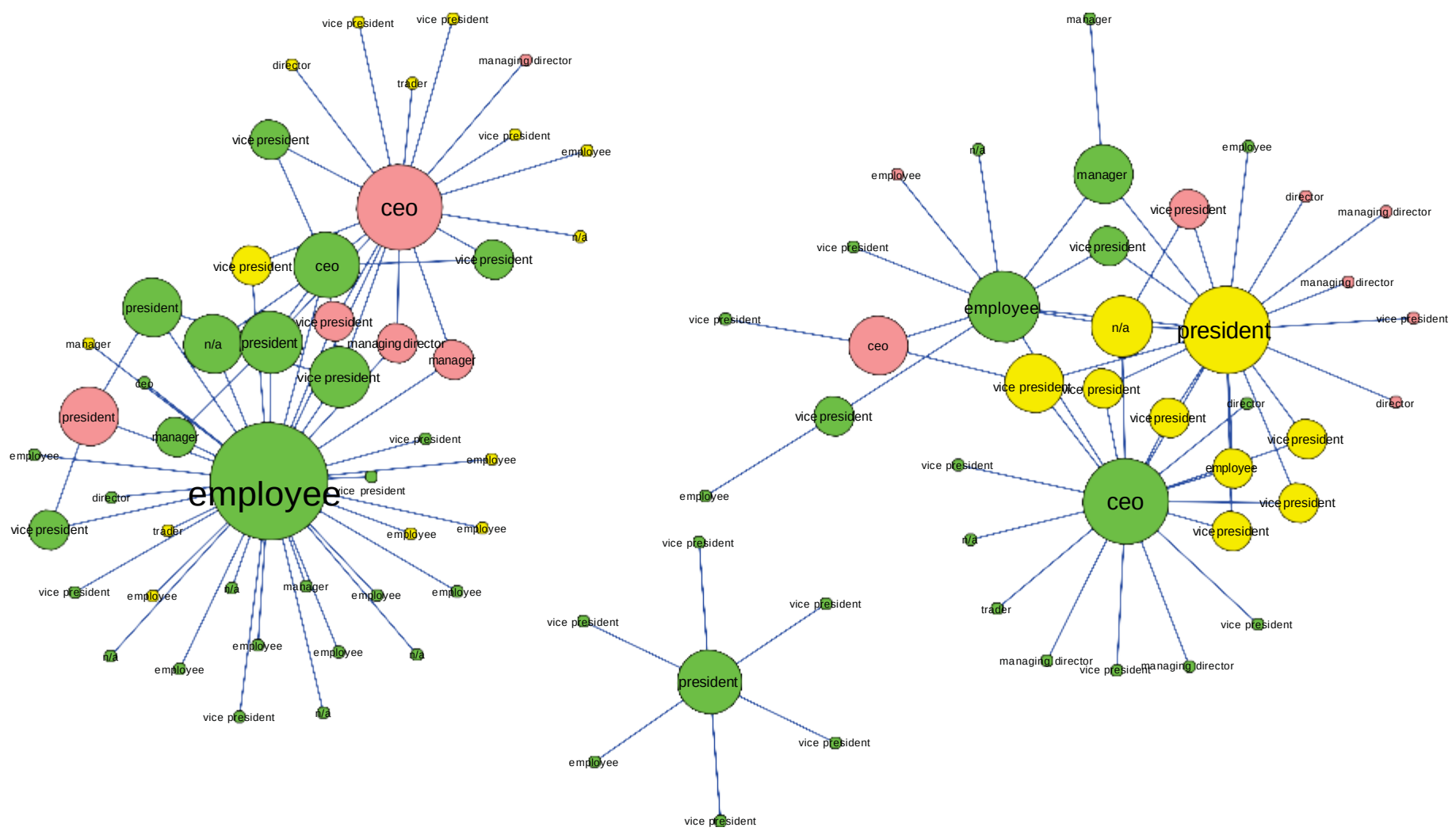

(a)

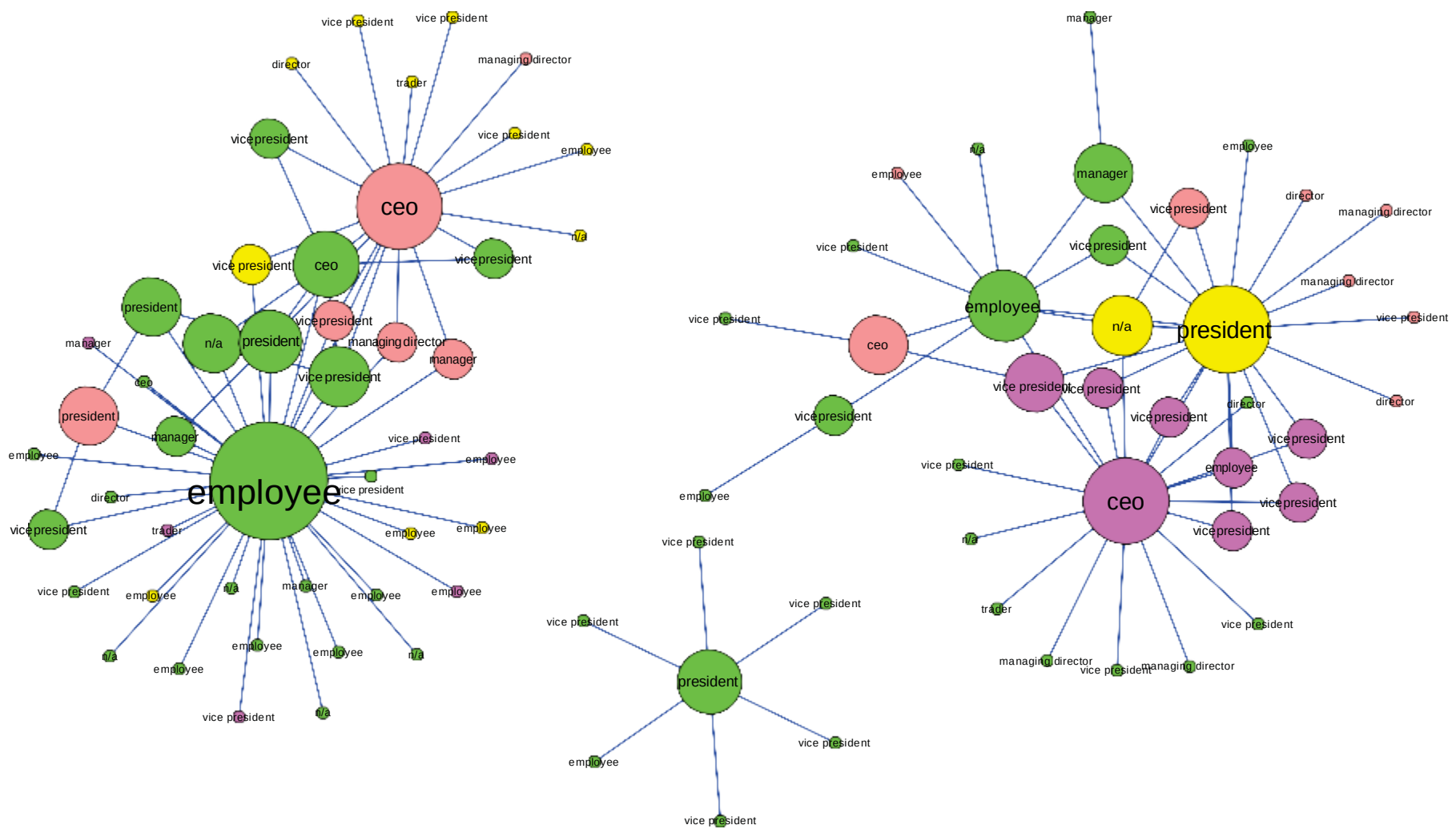

(b)

Figure 8: Sally Beck's Multi-account $\pi$-Net Communities, Clusters(with different colors): (a) Three (b) Four

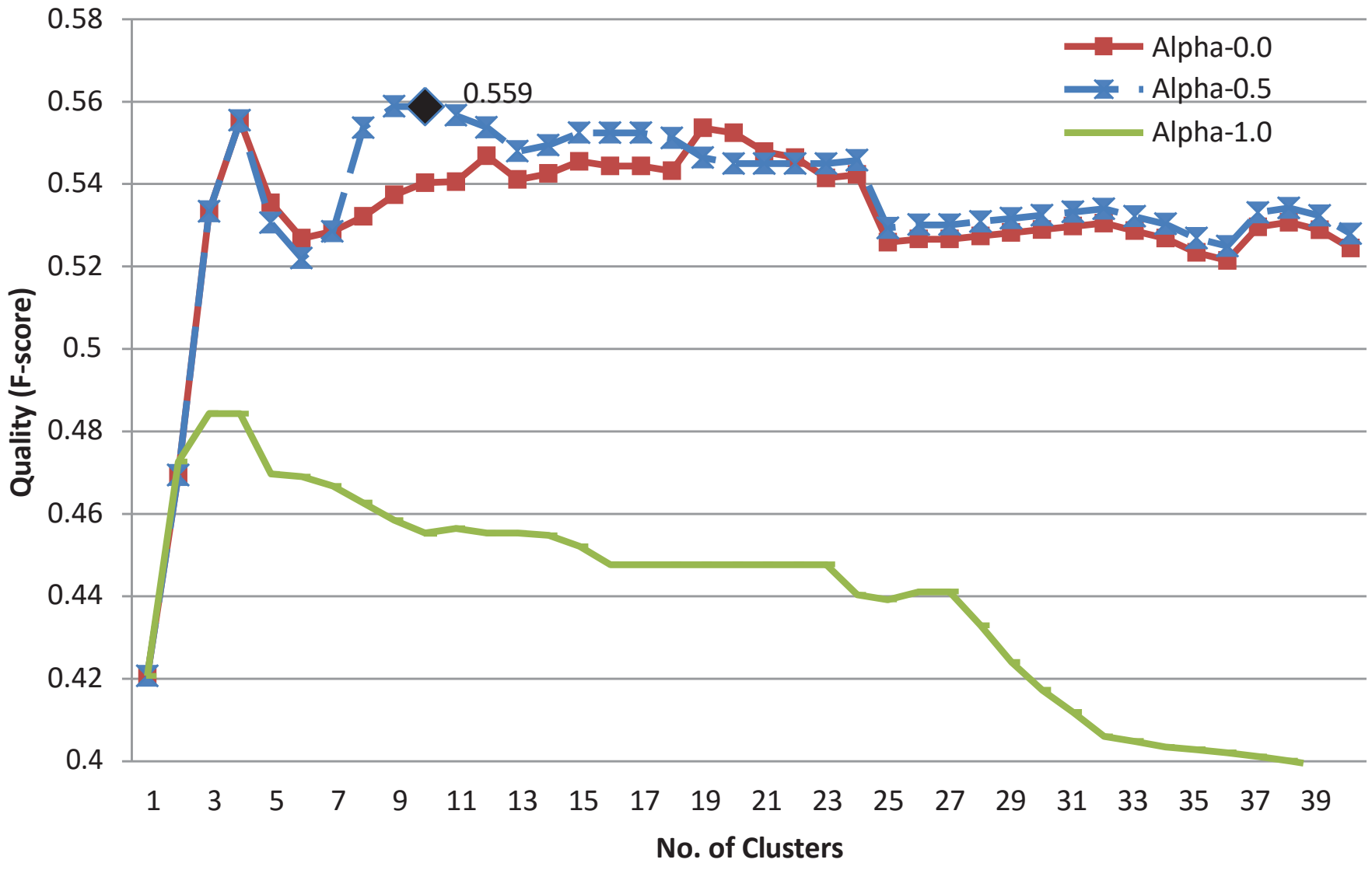

Figure 9: No. of Communities vs. F-Score (F-measure): Sally-Beck's Personalized Communities [single-account/fused $\pi$-Net]

In addition, the proposed idea is based on the shortest path strategy to compute the similarity among users that seems less effective than all paths approach. However, we can achieve approximately equivalent results in terms of density and entropy. In Fig. 10(a), SA approach(Cheng et al., 2011) is evaluated against proposed method in terms of structural and semantic measures. We analyze the results by incorporating information from single and multiple users to construct the $\pi$-Nets while the number of communities are fixed. The statistics of each $\pi$-Net is provided in Fig. 10(b). It is evident that the proposed method is capable of finding competitive results compared with existing method. The detailed analysis on this aspect can be found in (Nawaz et al., 2012).

### 5.3.2. *Community Dynamics*

We are interested to investigate the evolution of communities in terms of personalization. The community visualizations along with the statistical infor-mation seem a better way to understand the multi-user personalized approach.

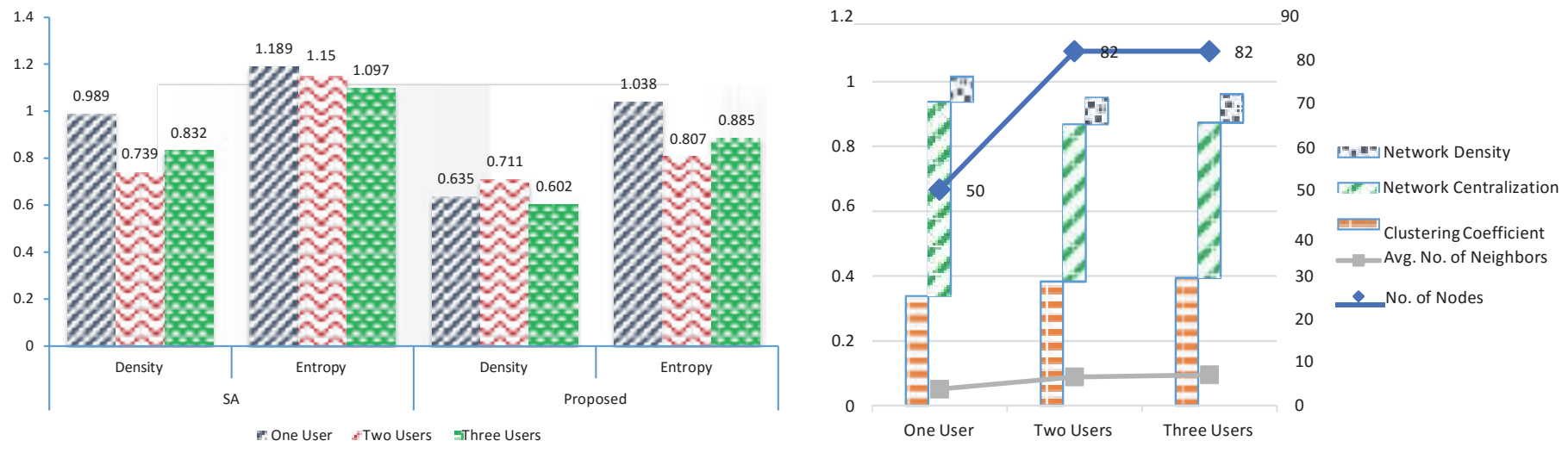

(a) (b)

Figure 10: Qualitative Analysis [(a)No. of Users vs. Quality, (b) $\pi$-Net Statistics], $\alpha$ = 0.5, K=2

Single-user personalized communities refer to disjoint groups of users, where each member of the groups has an exclusive email communications with others. However, members share emails among themselves through the reference user. In Fig.11 (a) Sally Beck's four personalized communities are presented. Every user is exactly two hops away from Sally Beck due to personalization constraint. Sally Beck, Louise Kitchen, John Lavorato, and Delainey David are the key members because they share emails with the majority members of each community. It is very interesting to figure out what happens if we relax the personalization constraint beyond single user. Furthermore, how the email community structure evolves as we include other user's information? How to identify and limit the potential users for effective multi-user personalized communities, which is beyond the scope of this paper.

With the intention of answering queries related to the dynamics of communities, we also contemplate the multi-user personalized communities, shown in Fig. 12(a), by randomly adding one of Sally Beck neighbor's information, i.e., Louise Kitchen. In this figure, the edge size reflects the strength of communication. The overall intensity of email sharing is high across the communities. It shows that users even with high intensity of email exchanges need not to be in the same community. Based on this fact we can say that users have diverse nature of communication at single frame of reference. For example, at office time, a user may share more emails with friends or family members rather than office colleague.

In comparison to single-user personalized communities in Fig.11 (a), each community retains their key members in Fig.12(a). This kind of retention property is strongly influenced by the choice of a new user. For example, if we choose a new user who have communicated with non-key member of the community through other members. In that case, non-key member have higher chances to be the key member. We also notice that extension of personalization concept may affect the membership of the users. For instance consider the community 3 in Fig.11 (a) where Jonathan Mckay, Mike Swerzbin, and Rogers Herndon

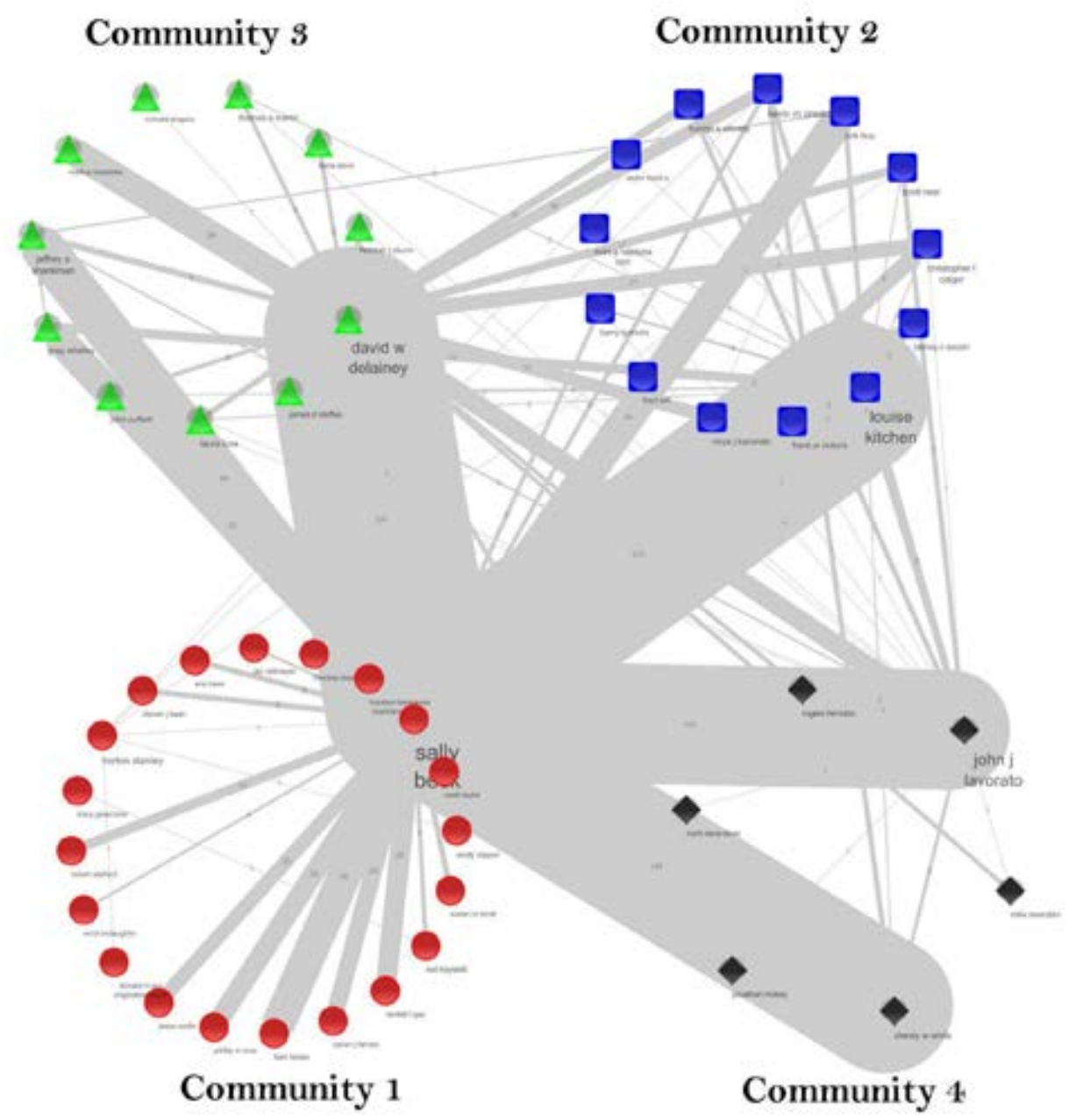

(a)

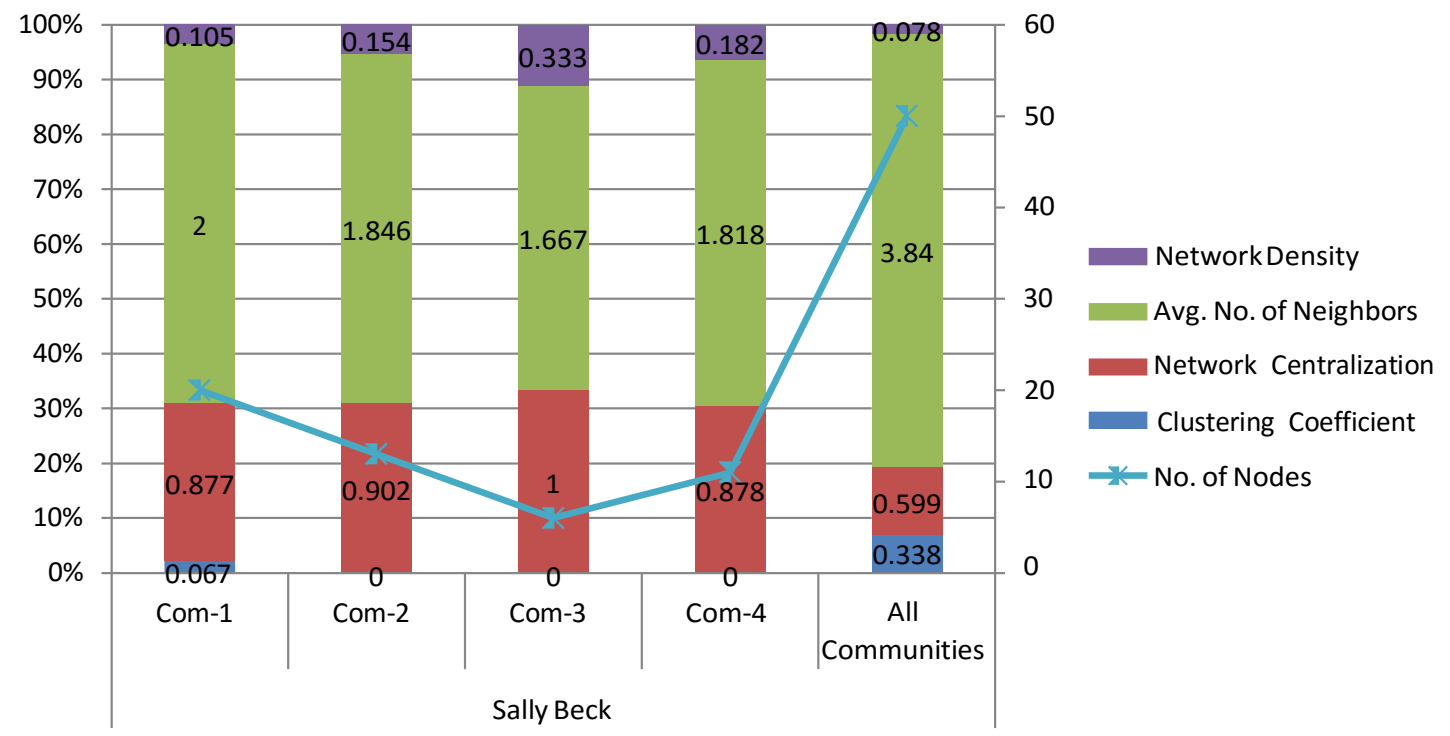

(b)

Figure 11: Single-User Personalized Communities using Sally-Beck's $\pi$-Net: (a) Four Communities (b) Community Statistics

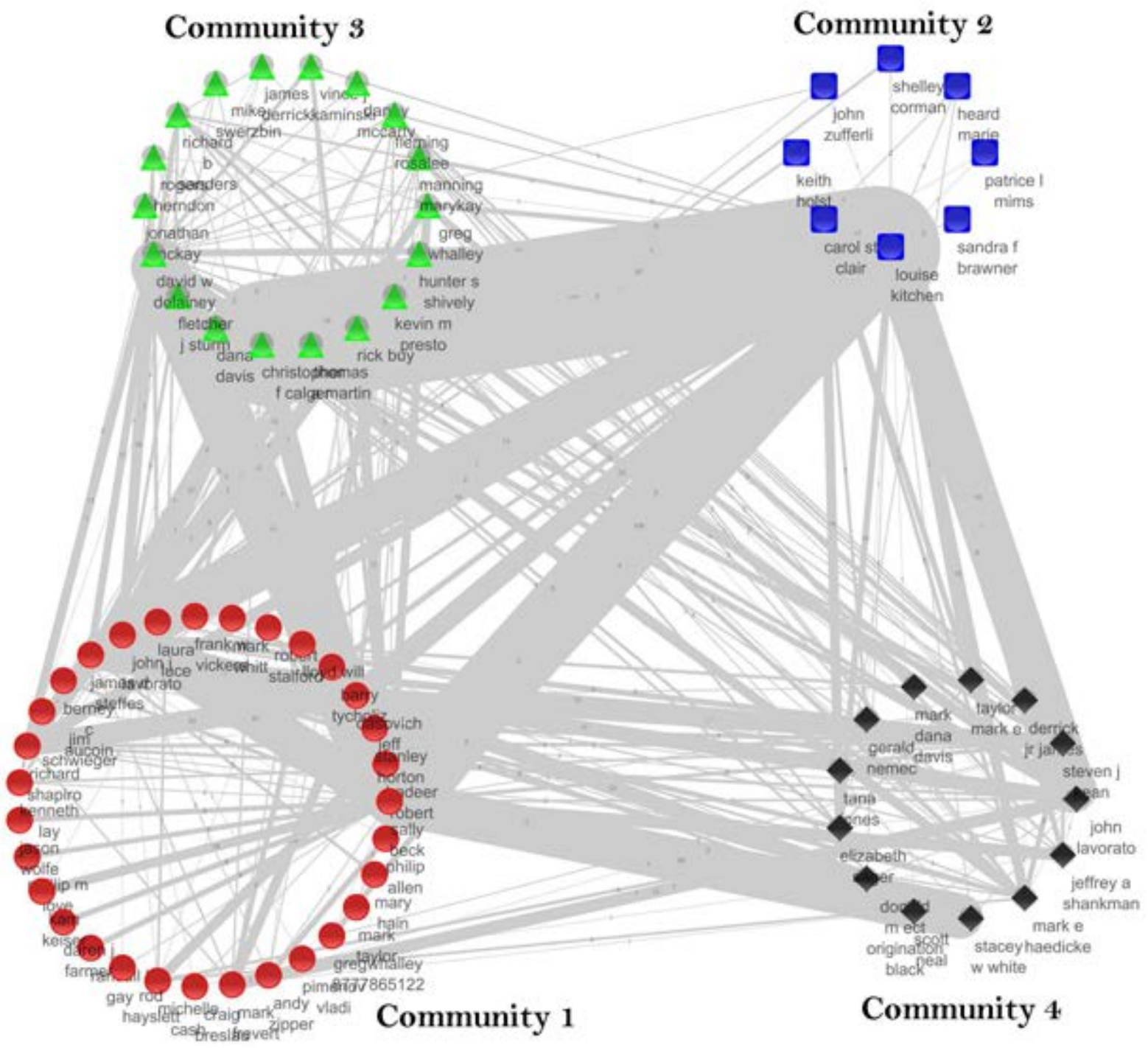

(a)

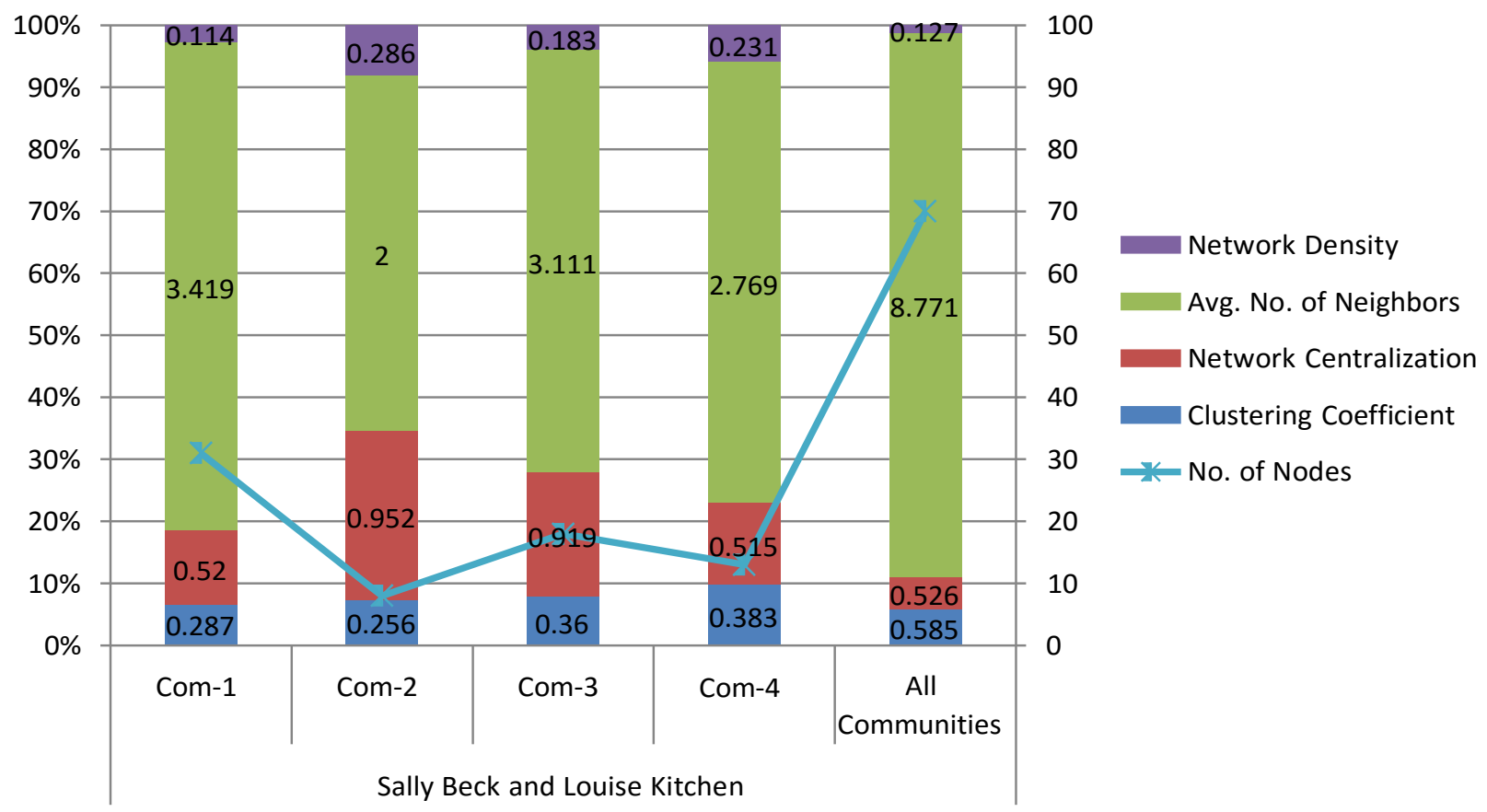

(b)

Figure 12: Multi-User Personalized Communities using Sally-Beck and Louise Kitchen's $\pi$-Nets: (a) Four Communities (b) Community Statistics

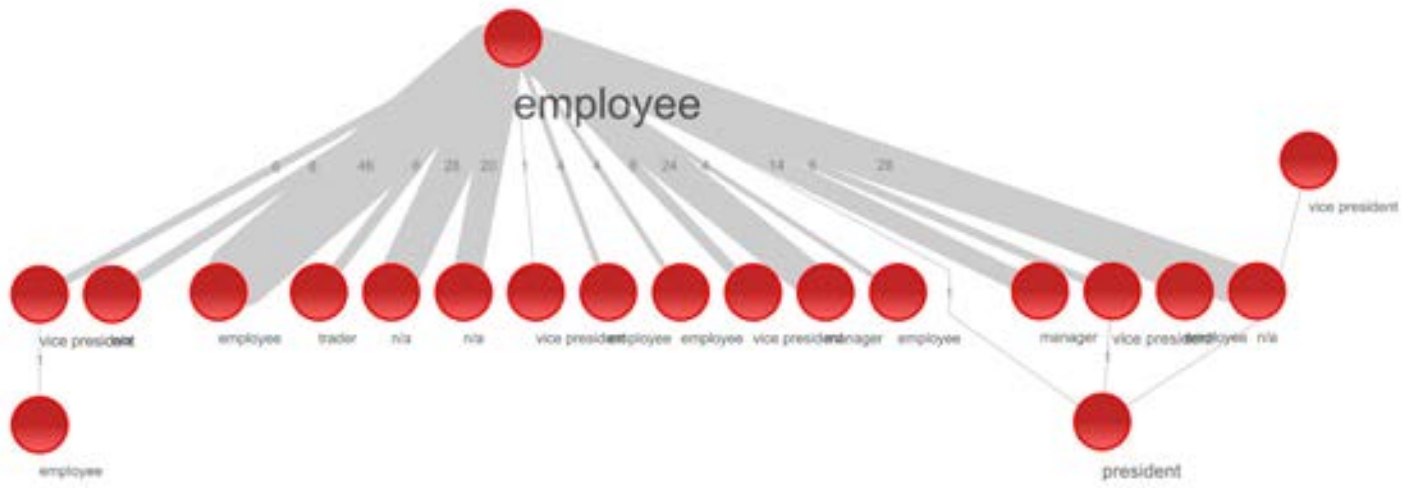

(a)

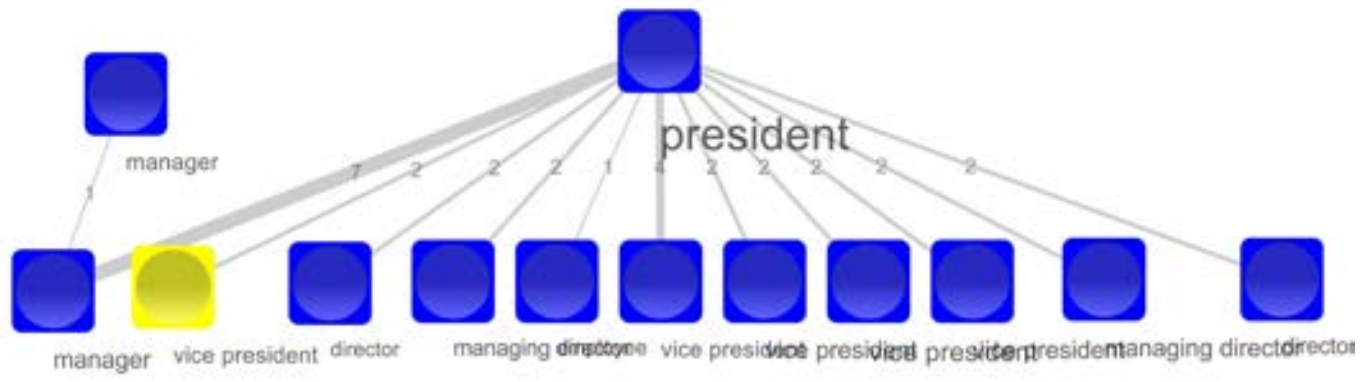

(b)

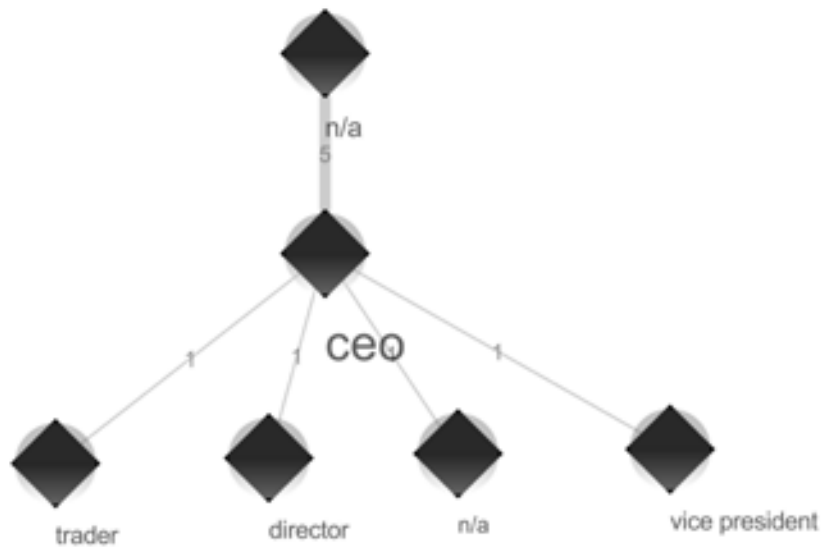

(c)

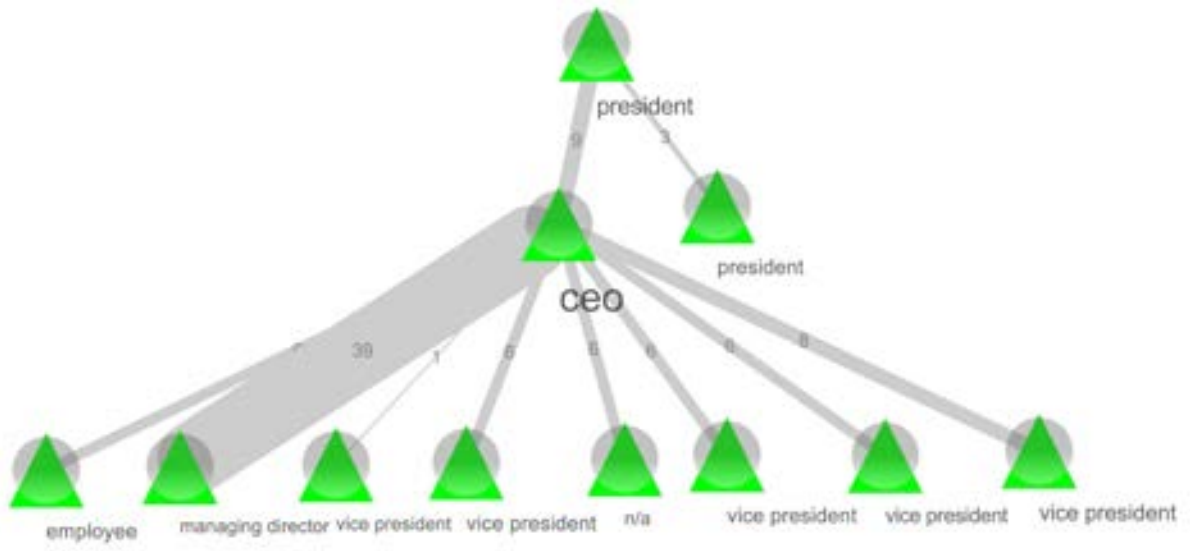

(d)

Figure 13: The Hierarchical View of Single-user $\pi$-Net Communities from Fig. 11: (a) Community 1 (b) Community 2 (c) Community 3 (d) Community 4

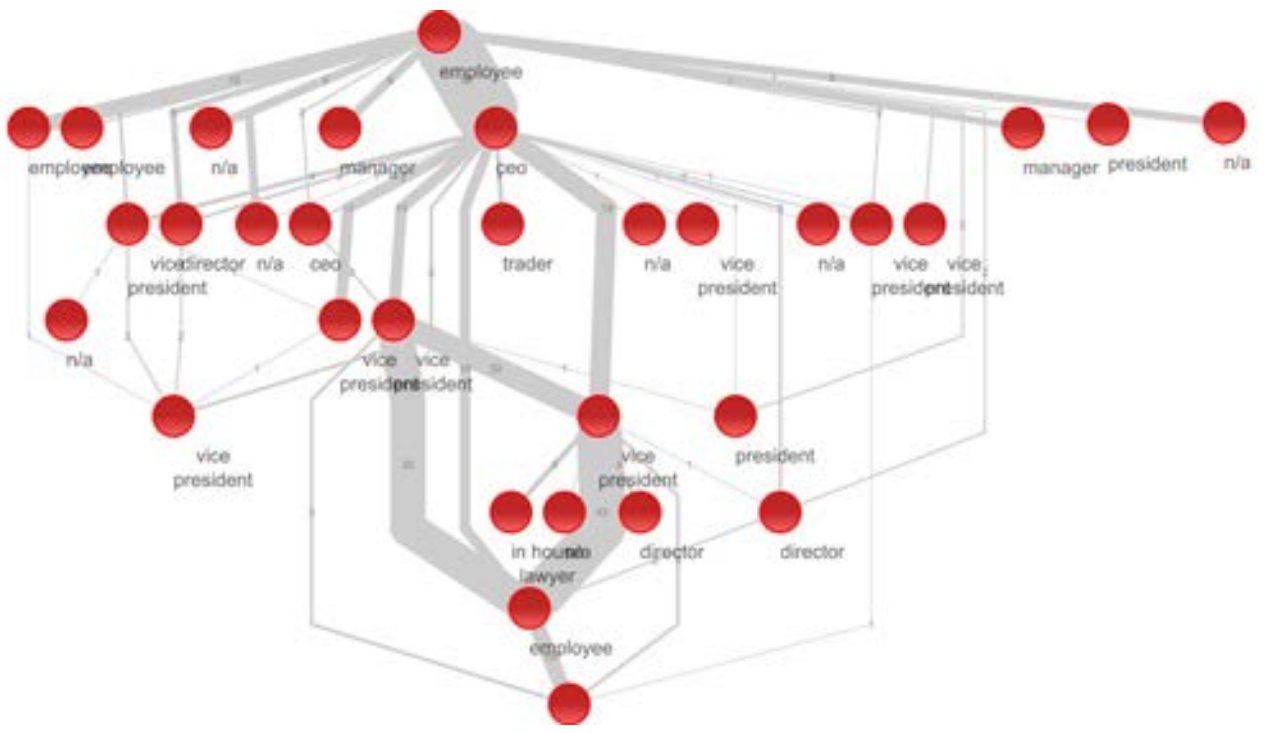

(a)

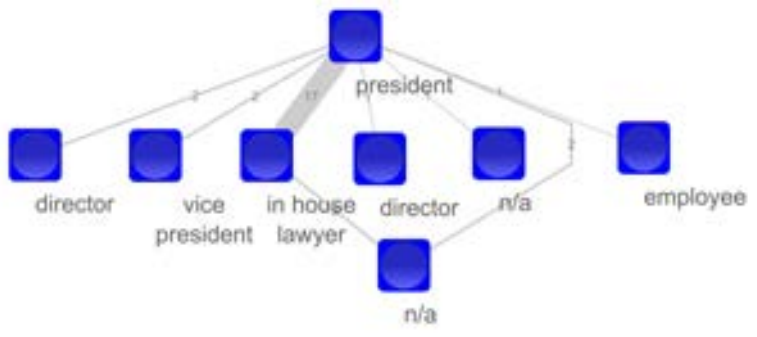

(b)

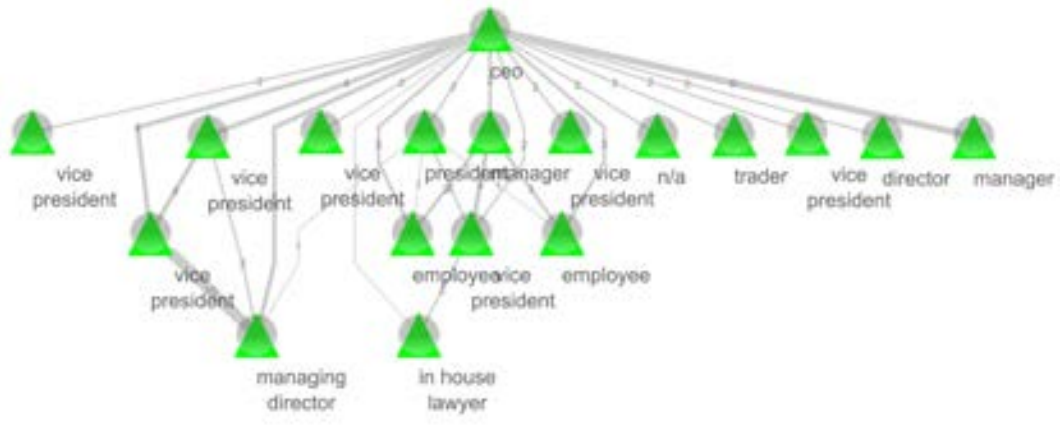

(c)

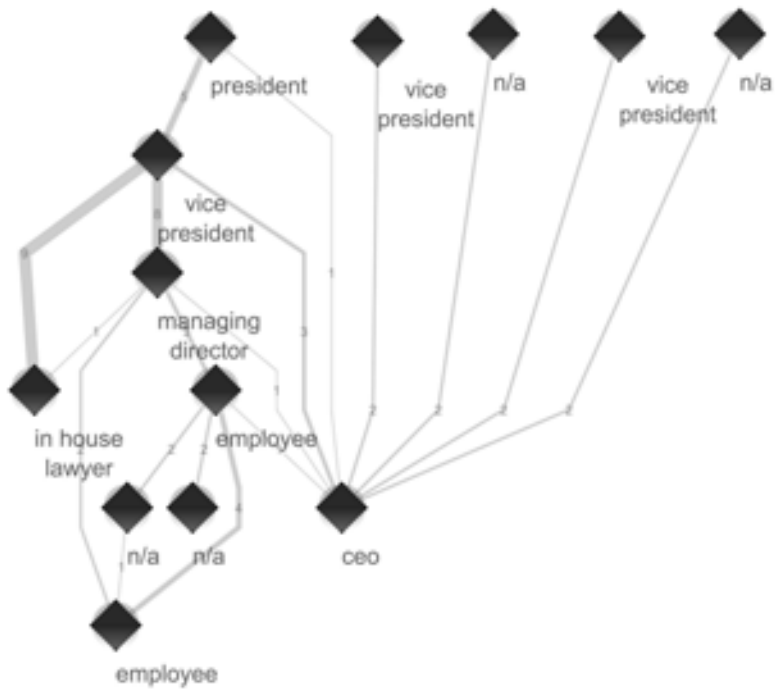

(d)

Figure 14: The Hierarchical View of Multi-user $\pi$-Net Communities from Fig. 12: (a) Community 1 (b) Community 2 (c) Community 3 (d) Community 4

become the members of community 4 in Fig.12(a).

The communication pattern analysis among users within a community is also thought-provoking idea especially in reference to hierarchical structure of the organization. Additionally, we do emphasize on community evolution process in this context. Fig.13 and Fig.14 provide the pictorial representation of the virtual (digital or email) communication in the physical world (designations of users). These communities do not reflect the homogeneous grouping of users with respect to their designations. Other way around, we can say that one person can have similar communication with heterogeneous hierarchical nature of the users. From this observation, we can infer that virtual interactions are more diverse and congested than actual communications. Furthermore, the evolution of communities does not change the nature of interactions rather the depth of interaction hierarchy, which we discuss in the subsequent section.

#### *5.3.3. Personalized Network Exploration*

The multi-user personalization concept can define an alternate way of network expansion. Moreover, a careful selection of users may allow us to approximate the network with restricted information. We analyze the impact of personalized expansion in terms of explored users and network properties.

The selection order of multiple users strongly influences the personalized network. We present the statistical view of personalized network evolution in Table 1. In this paper, we begin randomly with Sally Beck's account information through which we discover email communication with 93 other users, i.e., 58% users within Enron email network. Conceptually the next candidate for network expansion should have more disjoint users compared with Sally Beck, i.e., Louise Kitchen. We can reduce the search space up to 14% if we consider Louise Kitchen as the second user to extend the personalized network. It seems computation intensive task to identify such candidates. A heuristic approach could be to choose the next candidate across the communities to maximize the expansion.

It is easy to understand the dynamics of network structure by analyzing its parameters instead visual representation. The specific parameters[6] are calculated against each resultant community, as revealed in Fig. 11(b) and Fig. 12(b), to anticipate the topological changes in discovered communities. For example, the clustering coefficient, i.e., which defines the how strongly the neighbors of a particular vertex are connected, is comparatively higher in case of multi-user personalized communities. It is due to the fact of shared users between Sally Beck and Louise Kitchen. In contrast, the remaining parameters are quite stable for both scenarios. The communities discovered through multi-user personalization concept can be consumed in many applications areas. Though, few of them are highlighted in section 2.

**Managerial Insights:** The results of the proposed approach help to anticipate the evolution of communities, i.e. how email network properties are changed after incrementally adding new emails from other accounts. This kind of analysis helps the authorities to understand the nature of email communications and their relationships. For example, by analyzing their community dynamics, we can identify whether two email accounts are owned by one or two different people. Moreover, we can identify the lower bound of user number information to explore the entire email network. In other words, it is possible to approximate the entire email network structure by consuming few email accounts, since, all the users either in sender or recipient list contains the copy of corresponding email. It is also possible to analyze the hierarchical structure of authority and communication patterns, e.g. an employee in some organization is expected to frequently coordinate with his fellow employees or immediate boss rather than CEO.

[6]http://med.bioinf.mpi-inf.mpg.de/netanalyzer/index.php

Table 1: The evolution of network coverage using multi-user personalized information. The disjoint memebers for each $\pi$-Net are presented along with their designations in second column. Sally Beck & Louise Kitchen have communicated with 93 & 116 distinct users respectively. There are 161 distinct users in this dataset where 80% users of the entire network are exposed using two email accounts.

| Account Owner | Exclusive Members in $\pi$-Net | Network Coverage |
|---|---|---|
| Sally Beck | mclaughlin jr errol (employee), jay reitmeyer (employee), theresa staab (employee), horton stanley (president), eric bass (trader), mark davis (vice president) | Local(93/129, 72%), Global(93/161, 58%) |
| Louise Kitchen | benson robert (director), badeer robert (director), ermis frank (director), dasovich jeff (employee), manning marykay (employee), carson mike (employee), dorland chris (employee), semperger cara (employee), meyers albert (employee), salisbury holden (employee), solberg geir (employee), germany chris (employee), townsend judy (employee), phillip platter (employee), derrick jr james (in house lawyer), mary hain (in house lawyer), forney john m (manager), king jeff (manager), fischer mark (manager), teb lokey (manager), cuilla martin (manager), heard marie (n/a), thomas paul d (n/a), hernandez juan (n/a), philip allen (n/a), pimenov vladi (n/a), parks joe (n/a), campbell larry f (n/a), kim ward (n/a), stanley horton (president), ... | Local(116/129, 90%), Global(116/161, 72%) |

## 6. Related Work

In the recent past, many researchers have analyzed entire and partial email networks to study influential users, email prioritization, community structures, and organizational hierarchies based on statistical or structural analysis. We explore prominent approaches sequentially, as categorized in Fig. 1, with respect to their relevance to our work.

Among the early efforts in email classification based on statistical analysis, Steve Martin et al. (Martin et al., 2005) have highlighted the importance of outgoing email information among individuals rather than incoming emails. In this approach, they have suggested outgoing email communication for better understanding of an individual's behavior. They have also analyzed the contribution of the features extracted from emails to email classification. Conceptually, email features are constructed based on the presence of various objects like HTML, script tags, embedded images, hyper links, MIME types of file attachments etc. Moreover, a support vector machine (SVM) and Naïve Bayes classifiers have been employed to form a novel worm detection system. On the other hand, the characterization of emails and community identification is achieved by L. Johansen et al. (Johansen et al., 2007) which is based on the flow and frequency of email interaction among users. Accordingly, email frequency threshold was the basis for partitioning different users. It follows a simple rule, i.e., if two individuals have a large number of interactions with each other, then they have an association. However, the threshold value may vary for each user. It is very hard to generalize an email frequency for effective user grouping.

Few techniques are intended to predict the organizational structure solely based on interaction and relationship among individuals through emails. J. Shetty et al. (Shetty & Adibi, 2005) have considered three kinds of communication channels, including (a) emails (b) phone calls (c) meetings, to predict the individual role in that organization . Text mining and natural language processing is the basis of their approach, which is computationally intensive. An alternate method is also adopted by H. Yang et al. (Yang et al., 2010) for the same problem which sustains the computation cost. K. Yelupula et al. (Yelupula & Ramaswamy, 2008) and Anton Timofieiev et al. (Timofieiev et al., 2008) have introduced ranking strategies which are also popular for identifying influential users in the network. These techniques are inspired by sent & received emails or page rank. Unfortunately, this kind of analysis tends to determine the authority level of individuals based on their importance in the communication network rather than grouping centered on the nature of user interaction. These methods require an extensive access to entire email network.

Due to unavailability or ineffectiveness of the entire network information, few localized approaches have also been developed for community analysis (Hu & Lau, 2012) (Wu et al., 2012). On the other hand, Shinjae Yoo et al. (Yoo et al., 2009) have utilized the topological structure of the personalized email network, with certain features like (a) degree (b) clustering coefficient(c) betweenness centrality, for non-Spam Email prioritization. This method is limited to topological aspects. Similarly, a recent study, (Biswas & Biswas, 2015), is focused on ego centric community detection in network data and it is emphasized on structural aspect alone, i.e., reachability and isolability. Hanhe Lin (Lin, 2010) has identified hidden relationships in the email network to highlight mutual private communication. Usually, private relationships cannot lead us to a grouping of users with similar email usage patterns or behaviors.

There is a rich body of work in the email domain with regard to having diverse objectives. In contrast to existing approaches, our method discovers communities or groups of individuals based on similar email communication patterns and email network structures altogether. We determine the communication patterns among users through meta-data associated with the emails. Accordingly, we adapt the multi-attributed graph clustering strategy. Our approach consumes multi-user personalized emails instead of single or entire email network. Furthermore, we analyze the community dynamics and global view of communities using personalized email data.

## 7. Conclusion and Future Directions

This paper studied the notion of a multi-user personalized email network and communities to assist email systems. A uniquely attributed social graph modeling strategy is presented to reflect the email interactions among users. This social graph is clustered where users are grouped together based on their similar communication patterns. Multi-user communities have comparable density, but lower entropy values, using different clustering algorithms, compared with its counterpart. The community dynamics in terms of network properties helped the system to determine the relationship among users. For instance, Sally Beck and Louise Kitchen shared emails with a similar group of people and have an overlapped community structure, which proves their closeness. These user grouping results enable email systems to offer their customers useful and interesting applications such as an email strainer, dynamic group prediction, and fraudulent account detection.

The proposed strategy is based on meta-data, i.e. time-stamp, size, attachment existence and type, and subject length, extracted from user emails and did not process the email body for efficiency. The clustering algorithm has combined the structural and semantic similarity linearly and adapted a shortest path approach. In terms of quality, we achieved competitive results. The parameter value for number of communities is provided in advance as a random choice during experiments, since it was not the focal point.

The proposed method has the potential to be integrated with online email visualization tools, e.g. Immersion[7] initiated by MIT Media Lab, for personalized email community analysis. The Immersion tool also considers meta data from emails; therefore, we can extend it with better visual analysis. Moreover, the integration of such visualization tools in some third party applications, e.g. MS Outlook, Mozilla Thunderbird where multiple accounts are registered, is an interesting and favorable scenario for our approach. It is worthwhile to identify and utilize influential email accounts, instead of random selection, for effective email community analytics.

## Acknowledgement

This research was supported by the MSIP (Ministry of Science, ICT & Future Planning), Korea, under the ITRC (Information Technology Research Center) support program (NIPA-2014(H0301-14-1020)) supervised by the NIPA (National IT Industry Promotion Agency).

[7]https://immersion.media.mit.edu/